\input harvmac

\def\MM{{\cal M}}
\rightline{EFI-98-22, RI-4-98, IASSNS-HEP-98-52}
\Title{
\rightline{hep-th/9806194}
}
{\vbox{\centerline{Comments on String Theory on $AdS_3$}}}
\medskip

\centerline{\it
Amit Giveon${}^1$,  David Kutasov${}^{2,3}$,
Nathan Seiberg${}^4$}
\bigskip
\centerline{${}^1$Racah Institute of Physics, The Hebrew University,
Jerusalem 91904, Israel}
\centerline{${}^2$Department of Physics, University of Chicago,
5640 S. Ellis Av., Chicago, IL 60637, USA}
\centerline{${}^3$Department of Physics of Elementary Particles,
Weizmann Institute of Science, Rehovot, Israel}
\centerline{${}^4$School of Natural Sciences,
Institute for Advanced Study, Olden Lane, Princeton, NJ, USA}
\smallskip

\vglue .3cm
\bigskip

\noindent
We study string propagation on $AdS_3$ times a compact space from an
``old fashioned'' worldsheet point of view of perturbative string
theory.  We derive the spacetime CFT and its Virasoro and current
algebras, thus establishing the conjectured $AdS$/CFT correspondence
for this case in the full string theory.  Our results have
implications for the extreme IR limit of the $D1-D5$ system,
as well as to 2+1 dimensional BTZ black holes and their
Bekenstein-Hawking entropy.

\Date{6/98}

\def\journal#1&#2(#3){\unskip, \sl #1\ \bf #2 \rm(19#3) }
\def\andjournal#1&#2(#3){\sl #1~\bf #2 \rm (19#3) }

\def\hat{\widehat}
\def\ie{{\it i.e.}}
\def\eg{{\it e.g.}}

\def\etc{{\it etc}}

\def\tilde{\widetilde}

\def\frac#1#2{{#1\over#2}}

\def\half{\frac12}

\def\inbar{\,\vrule height1.5ex width.4pt depth0pt}
\def\IC{\relax\hbox{$\inbar\kern-.3em{\rm C}$}}
\def\IR{\relax{\rm I\kern-.18em R}}
\def\IP{\relax{\rm I\kern-.18em P}}

%
%
\def\np#1#2#3{Nucl. Phys. {\bf B#1} (#2) #3}
\def\pl#1#2#3{Phys. Lett. {\bf #1B} (#2) #3}

\def\prl#1#2#3{Phys. Rev. Lett. {\bf #1} (#2) #3}

\def\prd#1#2#3{Phys. Rev. {\bf D#1} (#2) #3}

\def\cmp#1#2#3{Comm. Math. Phys. {\bf #1} (#2) #3}
\def\cqg#1#2#3{Class. Quant. Grav. {\bf #1} (#2) #3}
\def\mpl#1#2#3{Mod. Phys. Lett. {\bf #1} (#2) #3}

\catcode`\@=11
\def\slash#1{\mathord{\mathpalette\c@ncel{#1}}}
\overfullrule=0pt

\def\HH{{\cal H}}

\def\LL{{\cal L}}
\def\NN{{\cal N}}

\def\PP{{\cal P}}

\def\SS{{\cal S}}

\def\underrel#1\over#2{\mathrel{\mathop{\kern\z@#1}\limits_{#2}}}

\catcode`\@=12


%

\def \sinh{{\rm sinh}}
\def \cosh{{\rm cosh}}

\def\exp{{\rm exp}}



\newsec{Introduction}

\lref\horwel{G. Horowitz and D. Welch, hep-th/9302126,
\prl{71}{1993}{328}.}
\lref\ks{D. Kutasov and N. Seiberg, \np{358}{1991}{600}.}
\lref\liouvilles{N. Seiberg,
in {\it
Common Trends in Mathematics and Quantum Field Theory (Proc. of the
1990 Yukawa International Seminar, Kyoto).} Prog.
Theor. Phys. Sup. {\bf 102}
319 (1990), edited by T. Eguchi, T. Inami and T. Miwa; in {\it
Random Surfaces and Quantum Gravity (Proc. of a Cargese meeting
1991)}, Plenum Press, New York and London (1991).}
The purpose of this paper is to study string propagation on curved
spacetime manifolds that include $AdS_3$. We will mostly discuss the
Euclidean version also known as $H_3^+= SL(2,C)/SU(2)$ (in Appendix A
we will comment on the Lorentzian signature version of $AdS_3$, which
is the $SL(2,R)$ group manifold).  At low energies the theory reduces
to $2+1$ dimensional gravity with a negative cosmological constant
coupled (in general) to a large collection of matter fields.  The low
energy action is
\eqn\lowener{\SS={1 \over 16 \pi l_p} \int d^3x
\sqrt{g}(R+{2 \over l^2}) + ...}
but we will go beyond this low energy approximation.

Our analysis has applications to some problems of recent interest:
\item{(a)} Brown and Henneaux
\ref\brhen{J. D. Brown and M. Henneaux, \cmp{104}{1986}{207}.}
have shown that any theory of gravity on $AdS_3$ has a large symmetry
group containing two commuting copies of the Virasoro algebra and thus
can presumably be thought of as a CFT in spacetime. The Virasoro
generators correspond to diffeomorphisms which do not vanish sufficiently
rapidly at infinity and, therefore, act on the physical Hilbert space.
In other words, although three dimensional gravity does not have local
degrees of freedom, it has non-trivial ``global degrees of freedom.''
We will identify them in string theory on $AdS_3$ as holomorphic (or
anti-holomorphic) vertex operators which are integrated over contours
on the worldsheet. Similar vertex operators exist in string theory 
in flat spacetime.  For example, for any spacetime gauge symmetry
there is a worldsheet current $j$ and $\oint j (z)$ is a good vertex
operator.  It measures the total charge (the global part of the gauge
symmetry).  The novelty here is the large number of such conserved
charges, and the fact that, as we will see, they can change the mass of
states.
\item{(b)} There is a well known construction of black hole solutions
in $2+1$ dimensional gravity with a negative cosmological constant
\lowener, known as the BTZ construction
\ref\btz{M. Banados, C. Teitelboim and J. Zanelli,
\prl{69}{1992}{1849}.}. BTZ black holes can be
described as solutions of string theory which
are orbifolds of more elementary string solutions
\ref\orbiele{M. Banados, M. Henneaux, C. Teitelboim and J. Zanelli,
\prd{48}{1993}{1506}; A. Steif, \prd{53}{1996}{5521}; D. Cangemi,
M. LeBlanc and R. Mann, \prd{48}{1993}{3606}.}. Strominger
\ref\strom{A. Strominger, hep-th/9712251.}
suggested a unified point of view for all black objects whose near
horizon geometry is $AdS_3$, including these BTZ black holes
and the black strings in six
dimensions discussed in
\ref\strvafa{A. Strominger and C. Vafa, hep-th/9601029,
\pl{379}{1996}{99}.},
and related their Bekenstein-Hawking entropy to the central charge
$c$ of the Virasoro algebra of \brhen. The states visible in
the low energy three dimensional gravity form a single 
representation of this Virasoro algebra. Their density of states
is controlled by \refs{\liouvilles, \ks} $c_{\rm eff}=1$, which
in general is much smaller than $c$. Our analysis shows that 
the full density of states of the theory is indeed controlled
by $c$ and originates from stringy degrees of freedom.    
\item{(c)}  Maldacena conjectured
\ref\malda{J. Maldacena, hep-th/9711200.}
(see
\ref\rkleb{S. S. Gubser, I. R. Klebanov, and A. W. Peet,
hep-th/9602135, Phys. Rev. {\bf D54} (1996) 3915;
I. R. Klebanov, hep-th/9702077, Nucl. Phys. {\bf B496} (1997) 231;
S. S. Gubser, I. R. Klebanov, A. A. Tseytlin, hep-th/9703040,
Nucl. Phys. {\bf B499} (1997) 217; S. S. Gubser and I. R. Klebanov,
hep-th/9708005, Phys. Lett. {\bf B413} (1997) 41.} 
for related earlier work and
\nref\gkp{S. Gubser, I. Klebanov and A. Polyakov, hep-th/9802109.}%
\nref\witt{E. Witten, hep-th/9802150.}%
\refs{\gkp,\witt} for a more precise statement of the conjecture)
that string theory on $AdS$ times a compact space is dual to a CFT.
Furthermore, by studying the geometry of anti-de-Sitter space 
Witten \witt\ 
argued on general grounds that the observables in a quantum theory of 
gravity on $AdS$ times a compact space should be interpreted as correlation
functions in a local CFT on the boundary.  Our work gives an explicit
realization of these ideas for the concrete example of strings on
$AdS_3$.  In particular, we construct the coordinates of the spacetime
CFT and some of its operators in terms of the worldsheet fields.
\item{(d)} For the special case of type IIB string theory on
\eqn\one{\MM=AdS_3 \times S^3 \times T^4}
Maldacena argued that it is equivalent to a certain two-dimensional
superconformal field theory (SCFT), corresponding to the IR limit of
the dynamics of parallel $D1$-branes and $D5$-branes (the $D1/D5$ system).
Our discussion proves this correspondence.
\item{(e)} In string theory  in flat spacetime integrated 
correlation functions on the worldsheet give S-matrix elements. In 
anti-de-Sitter spacetime there is no S-matrix.  Instead, the interesting
objects are correlation functions in the field theory on the boundary
\refs{\malda, \gkp, \witt}.  Although the spacetime objects of interest are
different in the two cases, we will see that they are computed by following
exactly the same worldsheet procedure. 
\item{(f)} Many questions in black hole physics and the
$AdS$/CFT correspondence circle around the concept of holography
\ref\holography{C. R. Stephens, G. 't Hooft and B. F. Whiting,
gr-qc/9310006; L. Susskind, hep-th/9409089.}.
Our analysis leads to an explicit identification of the boundary
coordinates in string theory.  We hope that it will lead to a better
understanding of holography.
\nref\gawedzki{K. Gawedzki, in Proceedings of NATO ASI Cargese 1991,
eds. J. Frohlich, G. 't Hooft, A. Jaffe, G. Mack, P.K. Mitter,
R. Stora. Plenum Press, (1992), hep-th/9110076.}%
\nref\teschner{J. Teschner, hep-th/9712256;  hep-th/9712258.}%
\nref\egp{J.M. Evans, M.R. Gaberdiel, M.J. Perry, hep-th/9806024.}

In section 2 we review the geometry of $AdS_3$ and consider the CFT
with this target space (for earlier discussions of this system see
\refs{\gawedzki - \egp} and references therein).  We then
show how the $SL(2)\times SL(2)$ current algebra on the string
worldsheet induces current algebras and Virasoro algebras in
spacetime.  This leads to a derivation of the $AdS$/CFT correspondence
in string theory.  In section 3 we extend the analysis to the
superstring, and describe the NS and R sectors of the
spacetime SCFT.
In section 4 we explain the relation between our system and
the dynamics of parallel strings and fivebranes. We discuss
both the case of $NS5$-branes with fundamental strings and the
$D1/D5$ system. We also
relate our system to BTZ black holes.  In
Appendix A we discuss the geometry of $AdS_3$ with Lorentzian
signature.  In Appendix B we discuss string theory on $\MM$ with
twisted supersymmetry.

\newsec{Bosonic Strings on $AdS_3$}

According to Brown and Henneaux \brhen, any theory of three
dimensional gravity with a negative cosmological constant has an
infinite symmetry group that includes two commuting Virasoro algebras
and thus describes a {\it two dimensional} conformal field theory in
spacetime. In this section we explain this observation in the context
of bosonic string theory on
\eqn\bosvac{AdS_3\times \NN}
where $\NN$ is some manifold (more generally, a target space for a
CFT) which together with $AdS_3$ provides a solution to the equations
of motion of string theory.

Of course, such vacua generically have tachyons in the spectrum, but
these are irrelevant for many of
the issues addressed here (at least up to a
certain point) and just as in many other situations in string theory,
once the technically simpler bosonic case is understood, it is not
difficult to generalize the discussion to the tachyon free
supersymmetric case (which we will do in the next section).

We start by reviewing the geometry of $AdS_3=H_3^+$.  It can be
thought of as the hypersurface
\eqn\defeq{-X_{-1}^2+X_3^2+X_1^2+X_2^2=-l^2}
embedded in flat $\CR^{1,3}$ with coordinates
$(X_{-1}, X_1, X_2, X_3)$. Equation \defeq\ describes a
space with constant negative
curvature $-1/l^2$, and $SL(2,C) \simeq Spin(1,3)$ isometry.
The space \defeq\ can be parametrized by the coordinates
\eqn\tthetacoo{\eqalign{
X_{-1}=&\sqrt{l^2+r^2}\cosh {\tau} \cr
X_3=&\sqrt{l^2+r^2}\sinh {\tau} \cr
X_{1}=&r\sin \theta \cr
X_{2}=&r\cos \theta \cr}}
(where $\theta \in [0,2\pi)$ and $r$ is non-negative) in terms of
which the metric takes the form
\eqn\metricttheta{ds^2=
\left(1+{r^2\over l^2}\right)^{-1}dr^2+
l^2 \left(1+{r^2\over l^2}\right)d\tau^2+r^2d\theta^2}
Another convenient set of coordinates is
\eqn\uycoo{\eqalign{
&\phi =  \log (X_{-1}+X_3)/ l \cr
&\gamma={X_2+iX_1 \over X_{-1}+X_3}\cr
&\bar \gamma={X_2-iX_1 \over X_{-1}+X_3}.\cr}}
Note that the complex coordinate $\bar \gamma$ is the complex
conjugate of $\gamma$. The surface \defeq\ has two disconnected
components, corresponding to $X_{-1}>0$ and $X_{-1}<0$. We will
restrict attention to the former, on which $X_{-1}>|X_3|$;
therefore, the first line of \uycoo\ is meaningful.
In the coordinates $(\phi,\gamma,\bar\gamma)$ the metric is
\eqn\metricuy{ds^2=l^2(d\phi^2 + e^{2\phi} d\gamma d\bar \gamma).}
The metrics \metricttheta\ and \metricuy\ describe the
same space. The change of variables between them is:
\eqn\changeva{\eqalign{
&\gamma={r \over \sqrt{l^2+r^2}} e^{-\tau+i\theta} \cr
&\bar \gamma={r \over \sqrt{l^2+r^2} }e^{-\tau-i\theta} \cr
&\phi =\tau+\half\log(1+{r^2 \over l^2} ). }}
The inverse change of variables is:
\eqn\inversecha{\eqalign{
&r=le^{\phi} \sqrt{\gamma \bar \gamma}\cr
&\tau=\phi - \half \log(1+e^{2\phi}\gamma\bar \gamma)\cr
&\theta={1 \over 2 i} \log (\gamma/\bar \gamma).\cr}}
It is important that both sets of coordinates cover the entire space
exactly once -- the change of variables between them \changeva\ and
\inversecha\ is one to one.

In the coordinates \metricttheta\ the boundary of
Euclidean $AdS_3$ corresponds to $r \rightarrow \infty$.
It is a cylinder parametrized by $(\tau,\theta)$.
The change of variables \changeva\ becomes for large $r$:
$e^\phi \approx re^\tau/l$, $\gamma \approx e^{-\tau+i\theta } $,
$\bar \gamma \approx e^{-\tau -i\theta}$. Thus, in the coordinates
\metricuy\ the boundary corresponds to $\phi\to\infty$;
it is a sphere parametrized by $(\gamma,\bar\gamma)$.

\subsec{Worldsheet Properties of Strings on $AdS_3$}

\lref\berfel{D. Bernard and G. Felder,
\cmp{127}{1990}{145}.}
\lref\berkut{M. Bershadsky and D. Kutasov, \pl{266} {1991}{345}.}
\lref\fms{D. Friedan, E. Martinec and S. Shenker,
\np{271}{1986}{93}.}

To describe strings propagating on the space \defeq\ we need to add a
(Neveu-Schwarz) $B_{\mu\nu}$ field in order to satisfy the equations
of motion. {}From the worldsheet point of view this is necessary for
conformal invariance.  The necessary $B$ field is $B= l^2 e^{2\phi}
d\gamma \wedge d \bar \gamma$.  Note that it is imaginary.  Therefore,
the worldsheet theory is not unitary. With a Euclidean worldsheet 
the contribution of the $B$ field to the action is real and the 
theory is not reflection positive.  In this respect our
system is different from the analytic continuation to flat Euclidean
space of strings in flat Minkowski space.  The worldsheet Lagrangian
with the $B$ field is
\eqn\wslag{\LL = {2l^2\over l_s^2} \left( \partial \phi \bar\partial
\phi + e^{2\phi} \bar \partial \gamma \partial \bar \gamma\right) }
($l_s$ is the fundamental string length). Note that with a Euclidean
signature worldsheet $\LL$ is real and bounded from below; therefore,
the path integral is well defined\foot{This is one of the reasons we
limit ourselves to the Euclidean problem of strings on $H_3^+$.  Had
we worked with a Lorentzian signature target space (the $SL(2,R)$
group manifold), the Euclidean worldsheet action would have had a real
part which is not bounded from below and the path integral would have
been ill defined.}.  Some of the $SL(2)$ symmetry is manifest in the
Lagrangian \wslag; \eg\ we can shift $\gamma $ by a holomorphic
function. It is convenient to add a one form field $\beta$ with spin
$(0,1)$ and its complex conjugate $\bar \beta$ with spin $(1,0)$, and
consider the Lagrangian
\eqn\wslagb{\LL={2l^2\over l_s^2}\left(  \partial \phi \bar\partial
\phi + \beta \bar \partial \gamma  + \bar\beta  \partial \bar\gamma -
e^{-2\phi} \beta \bar\beta \right).}
Integrating out $\beta$ and $\bar\beta$ we recover \wslag.
As in Liouville theory, at the quantum level the exponent in the last
term is renormalized.  Similarly, a careful analysis of the measure
shows that a dilaton linear in $\phi$ is generated. Taking these
effects into account and rescaling the fields one finds the worldsheet
Lagrangian
\eqn\acwak{\LL=\partial\phi\bar\partial\phi-
{2\over\alpha_+}\hat R^{(2)}\phi+\beta\bar\partial
\gamma+\bar\beta\partial\bar\gamma-\beta\bar\beta\exp\left(
-{2\over\alpha_+}\phi\right)}
where $\alpha_+^2=2k-4$ is related to $l$, the radius of
curvature of the space \defeq, via:
\eqn\tb{l^2=l_s^2 k.}
The Lagrangian \acwak\ leads to the free field
representation of $SL(2)$ current algebra
\ref\wakim{M. Wakimoto, \cmp{104}{1986}{605}.}
(see also \refs{\berfel,\berkut}).
It uses a free field $\phi$ and a holomorphic bosonic
$\beta,\gamma$ system \fms\ (as well as its anti-holomorphic
analog $\bar\beta,
\bar\gamma$) with weights $h(\beta)=1$,
$h(\gamma)=0$.
The last term in $\LL$ \acwak\ can be thought
of as a screening charge.
Correlation functions in the CFT
that are dominated by the region $\phi\to\infty$ (such
as bulk correlation functions \berkut) can be studied
by perturbing in this term; this leads to a prescription
similar to that used in Liouville theory. Generic correlation
functions are non-perturbative in the screening charge.

We can repeat a similar analysis in the $r,\theta, \tau$ variables.
After introducing new fields $\alpha$ and $\bar \alpha$ the Lagrangian
for large $r$ becomes
\eqn\acwaka{\LL={1 \over r^2} \partial r \bar\partial r +\alpha
\bar\partial (\tau-i\theta) +\bar\alpha \partial(\tau+ i \theta).}
In this limit $\log r$ is a free field which is a sum of a holomorphic
and an anti-holomorphic field.  Similarly, $\tau$ and $\theta$ are free
fields with holomorphic and anti-holomorphic components.  However, the
equations of motion also guarantee that $\tau-i\theta$ is
holomorphic.  This is consistent with the fact that for large $r$ it
is related to the holomorphic field $\gamma \approx
e^{-\tau+i\theta}$ (see \changeva).

A related description of CFT on Euclidean $AdS_3$ is obtained by
constructing the worldsheet Lagrangian using the $r,\theta, \tau$
coordinates and performing a T-duality transformation on $\theta$
\horwel.  In terms of the dual coordinate $\tilde \theta$ there is no
$B$ field; instead there is a dilaton field which is linear in $\log
r$. The Lagrangian is
\eqn\eqtdul{\LL= \partial \tau \bar \partial \tau + {1 \over r^2
}\partial \tilde \theta \bar \partial \tilde \theta +  {1 \over r^2+1}
\partial r \bar \partial r - 2i\partial \tilde \theta \bar \partial
\tau .}
Note that it has an imaginary term reflecting the lack of unitarity of
the system.  In terms of $\hat \tau = \tau-i\tilde \theta$ it is
\eqn\eqtduls{\partial \hat \tau \bar \partial \hat \tau + {r^2+ 1
\over r^2 }\partial \tilde \theta \bar \partial \tilde \theta +  {1
\over r^2+1} \partial r \bar \partial r.}
This description of the theory is similar but not identical
to that of \wslagb, \acwak,
\acwaka.  For large $r$ the theory becomes free and the corrections to
free field theory can be treated as a screening charge.

The theory has an affine $SL(2,R)\times SL(2,R)$ Lie algebra symmetry
at level $k$, generated by worldsheet currents $J^A(z)$, $\bar
J^A(\bar z)$, which satisfy the OPE:
\eqn\two{J^A(z)J^B(w)= {k\eta^{AB}/2
\over(z-w)^2}+{i\eta_{CD}\epsilon^{ABC}J^D\over z-w}+\cdots,
\qquad A,B,C,D=1,2,3}
where $\eta^{AB}$ is the metric on $SL(2,R)$ (with signature $(+,+,-)$)
and $\epsilon_{ABC}$ are the structure constants of $SL(2,R)$.  A
similar formula describes the operator products of the worldsheet
currents with the other chirality, $\bar J^A(\bar z)=(J^A(z))^*$.  The
level of the affine Lie algebra, $k$, is related to the cosmological
constant via eq. \tb.  The central charge of this model is:
\eqn\cntrch{c={3k\over k-2}.}

It will be useful for our purposes to recall the free field \acwak\
realization of $SL(2,R)$ current algebra \wakim. The worldsheet
propagators that follow from \acwak\ are:
$\langle\phi(z)\phi(0)\rangle=-\log|z|^2$,
$\langle\beta(z)\gamma(0)\rangle=1/z$.  The current algebra is
represented by (normal ordering is implied):
\eqn\curalg{\eqalign{
J^3=&\beta\gamma+{\alpha_+\over2}\partial\phi\cr
J^+=&\beta\gamma^2+\alpha_+\gamma\partial\phi
+k\partial\gamma\cr
J^-=&\beta.\cr}}
Interesting vertex operators are
\eqn\primr{V_{jm\bar m }=\gamma^{j+m}\bar \gamma^{j+ \bar m}
\exp\left( {2j\over\alpha_+}\phi\right).}
The exponents of $\gamma$ and $\bar \gamma$ can be both positive and
negative.  The only constraint that follows from single valuedness on
$AdS_3$ is that $m-\bar m$ must be an integer.  Obviously, $m-\bar m$
is the momentum in the $\theta$ direction.  One can check that $j$,
$m$ and $\bar m$ are the values of the $j$ quantum number of $SL(2,R)$,
and the $J^3$ and $\bar J^3$ quantum numbers, respectively\foot{Our
group is really the infinite multiple cover of $SL(2,R)$ (see 
Appendix A) and therefore $j,m,\bar m$ are not restricted to be 
half integers.}. The scaling dimension of $V_{jm\bar m}$ is 
$h=-j(j+1)/(k-2)$.

Which $SL(2,R)$ representations should we consider?  The affine
$SL(2,R)$ algebra does not have unitary representations.  This 
should not bother us because, as we said above, our worldsheet theory
is not unitary.  The problem that we are interested in is string
theory and therefore we should use the $SL(2,R)$ representations which
lead to a unitary string spectrum.  One way to do this is the following.
Consider the affine $U(1) \subset SL(2,R)$ generated by $J^3$ (the
``timelike direction'') and decompose each $SL(2,R)$ representation in
terms of the coset $SL(2,R)/U(1)$ and the $U(1)$ representation.  In
constructing a string vacuum we need the $SL(2,R)/U(1)$ coset to be
unitary.  The conditions for that were analyzed in
\ref\dpl{L. J. Dixon, M. E. Peskin and J. Lykken, \np{325}{1989}{329}.} 
with the conclusion
\eqn\unitaritycon{-1 < j <{k\over 2} -1, \qquad\qquad\qquad 2<k.}
Imposing the constraint \unitaritycon\ in string theory gives
rise to a unitary theory (see \eg\ \egp\ and references therein).

\subsec{Spacetime Properties of Strings on $AdS_3$}

Due to the presence of the worldsheet affine
$SL(2)$ Lie algebra \two\ the spacetime
theory has three conserved charges
\eqn\spcons{\eqalign{
L_0=&-\oint dz J^3(z)\cr
L_1=&-\oint dz J^+(z)\cr
L_{-1}=&-\oint dz J^-(z)\cr
}}
which satisfy the $SL(2,R)$ algebra $[L_n,L_m]=(n-m)L_{n+m}$
$(n,m=0,\pm1)$. The observations of \brhen\ lead one
to expect that \spcons\ should be extended to an
infinite dimensional Virasoro algebra with
central charge:
\eqn\virfull{[L_n,L_m]=(n-m)L_{n+m}+{c\over12}(n^3-n)\delta_{n+m,0}}
Our next task is to derive \virfull\ in string theory
and compute the central charge $c$.

As a warmup exercise, consider the following
related problem. Take the worldsheet CFT on
the manifold $\NN$ to contain an
affine Lie algebra $\hat G$ for a compact group $G$,
generated by currents $K^a$ satisfying the OPE:
\eqn\glss{K^a(z)K^b(w)=
{k'\delta^{ab}/2
\over(z-w)^2}+{if^{ab}_{\,\,\,\,\,\,c}K^c\over z-w}+\cdots;
\qquad a,b,c=1,\cdots, {\rm dim}\; G}
with $k'$ the level of $\hat G$.
Normally, this leads to the existence in the
spacetime theory of ${\rm dim}\, G$ conserved charges
\eqn\tkcons{T_0^a=\oint dz K^a(z)}
satisfying the algebra
\eqn\algt{[T^a_0,T^b_0]=if^{ab}_{\,\,\,\,\,\,c}T^c_0.}
However, in our case the spacetime theory
is a two dimensional CFT and we expect the charges
$T^a_0$ to correspond to the zero modes of an infinite
symmetry -- an affine Lie algebra {\it in spacetime},
generated by charges $T^a_n$ satisfying the
commutation relations
\eqn\kacmd{\eqalign{
[T^a_n,T^b_m]=&if^{ab}_{\,\,\,\,\,\,c}T^c_{n+m}+{\tilde k\over2}
n\delta^{ab}\delta_{n+m,0}\cr
[L_m,T^a_n]=&-nT^a_{n+m}\cr
}}
where $\tilde k$ is the level of affine $\hat G$
in spacetime.
We will next construct the operators $T^a_n$,
verify the first line of \kacmd, and compute
$\tilde k$. Later, when we
define the $\{L_m\}$ we will also verify the
second line.

The second line of \kacmd\ with $m=0$ states that the operators
$T^a_n$ carry $-n$ units of $L_0$ or $n$ units of $J^3$ \spcons. Thus,
in order to construct them we need to generalize the definition
\tkcons\ by multiplying the integrand $K^a(z)$ by a vertex operator
that carries $J^3$ but has worldsheet
scaling dimension zero and is holomorphic, so that it can be
integrated over $z$.  There is a unique candidate, the field
\primr\ $V_{j=0, m, \bar m=0}=\gamma^m$ with integer $m$.  Thus, we
define
\eqn\genta{T^a_n=\oint dz K^a(z)\gamma^n(z)}
and compute the commutator using standard techniques:
\eqn\commuta{[T^a_n, T^b_m]=
\oint dw\oint dz K^a(z)\gamma^n(z) K^b(w)\gamma^m(w)}
where the integral over $z$ is taken as usual along a small contour
around $w$, and the integral over $w$ is taken around some origin $0$.
The only source of singularities in the contour integral of $z$ around
$w$ comes from the OPE of currents \glss\ (the OPE of $\gamma$'s is
regular). The second term in the OPE \glss\ gives a first order
pole that is easily integrated to give:
\eqn\secter{if_{abc}\oint dw\oint dz K^c(w)\gamma^{n+m}(w)
{1\over z-w}=if_{abc}\oint dw K^c \gamma^{n+m}=if_{abc} T^c_{n+m}}
The first term in \glss\ gives a second order pole and needs
to be dealt with separately:
\eqn\sscctt{{k'\delta^{ab}\over2}\oint dw\oint dz
{\gamma^n(z)\gamma^m(w)\over (z-w)^2}
={k'\delta^{ab}\over2}\oint dw \partial_w(
\gamma^n)\gamma^m={nk'\delta^{ab}\over2}
\oint dw \gamma^{n+m-1}\partial_w\gamma
}
The r.h.s. of \sscctt\ is central --
it commutes with the generators $T^a_n$, \genta,
and more generally with all physical vertex operators in
the theory.  Therefore, this charge is not carried by
the excitations of the string but only by the vacuum.
The charge is non-vanishing only
for $n+m=0$ because otherwise $\oint dw \gamma^{m+n-1} \partial_w
\gamma = { 1\over m+n} \oint dw \partial_w \gamma^{m+n}=0$.  For
$n+m=0$ the integral
\eqn\pdef{p\equiv \oint dz{\partial_z\gamma \over \gamma}}
can be nonzero.  It counts the number of
times $\gamma$ winds around the origin when $z$
winds once around $z=0$.  Since $\gamma$ is a single
valued function of $z$, $p$ must be an integer\foot{For
Lorentzian signature $\gamma$ is real (see Appendix A),
and there is no natural definition of winding. This is another
reason for studying the Euclidean version of the theory.}.

To understand the meaning of the integer $p$, recall the spacetime
interpretation of the free field $\gamma$, \changeva.  In the
spacetime CFT at $r\to\infty$ \changeva, $\gamma=e^{-\tau+i\theta}$ is
a coordinate on the sphere with two punctures (corresponding
to $\tau=\pm\infty$).  Therefore, $p$ measures the number of
times the string worldsheet wraps around $\theta$.  We interpret
string theory on $AdS_3$ as having $p$ stretched fundamental strings
at $r\to\infty$.  The excitations of the vacuum described by vertex
operators correspond to small fluctuations of these infinitely
stretched strings and this is the reason they do not carry the
charge \pdef. String vacua with different values of $p$ correspond to
different sectors of the theory.

It is important that our target space is simply connected.  Therefore,
there cannot be any winding perturbative string states and hence $p$
commutes with all vertex operators describing
perturbative states.

Collecting all the terms \secter, \sscctt\ we
find that the $T^a_n$ satisfy the algebra \kacmd,
with
\eqn\stlevel{k_{\rm spacetime}\equiv \tilde k=pk'}
Thus, the affine Lie algebra
structure is lifted from the worldsheet to spacetime
and the level of the affine Lie algebra in spacetime
is equal to $p$ times that on the worldsheet.

\nref\green{M. B. Green, \np{293}{1987}{593}.}%
\nref\elie{S. Elitzur, A. Forge and E. Rabinovici, 
\np{388}{1992}{131}.}%
\nref\km{D. Kutasov and E. Martinec, hep-th/9602049,
\np{477}{1996}{652}.}%
\nref\gsw{M. Green, J. Schwarz and E. Witten,
{\it Superstring Theory}, Vol. 1, section 2.3.2.}%
A few comments are in order here:
\item{(a)} The fact that $p$ is a positive integer
is important to get a unitary realization of $\hat G$
in spacetime.
\item{(b)} We see that in string theory on $AdS_3$ there
is a close correspondence between worldsheet and spacetime
properties. A left-moving affine Lie algebra on the
worldsheet gives rise to a left-moving affine Lie algebra
in spacetime, \etc. This correspondence,
seen here and in many other aspects of our analysis below,
is reminiscent of analogous phenomena in theories
of worldsheets for worldsheets and related ideas \refs{\green - \km}.
\item{(c)} The derivation and, in particular, the treatment of
the integral in \sscctt\ makes it clear that one
should think of $\gamma$ as a holomorphic coordinate
in spacetime, in agreement with the geometric analysis
of eqs. \metricuy\ -- \inversecha. Note that $\gamma$ depends
holomorphically on the worldsheet coordinates, $\bar\partial
\gamma=0$. This is another example of the worldsheet -- spacetime
connection mentioned in item (b).
\item{(d)} The discussion above is
very reminiscent of the construction of DDF states
in string theory (see \gsw\ for details).
One can think of the operators $T^a_n$ \genta\
as a spectrum generating algebra.

\noindent
We are now ready to turn to the original
problem of finding the spacetime Virasoro
algebra. We proceed in complete analogy
with \genta\ -- \sscctt\
but there are a few new elements.
On general grounds we expect the Virasoro
generators to be given by
\eqn\virgen{L_n=\oint dz\left(a_3 J^3\gamma^n+
a_-J^-\gamma^{n+1}+a_+J^+\gamma^{n-1}\right)}
The operators \virgen\ are very similar to
photon vertex operators in a three
dimensional curved space. As usual, only one of
the three polarizations in \virgen\ is physical.
First, we have to impose BRST invariance, \ie\ require
the operator in brackets to be primary under the
worldsheet Virasoro algebra. This gives rise to
the constraint
\eqn\brsinv{na_3+(n+1)a_-+(n-1)a_+=0}
Furthermore, the fact that ``longitudinal photons''
are BRST exact and decouple leads to the identification
\eqn\longphot{(a_3, a_-, a_+)\simeq (a_3, a_-, a_+)+
\alpha(1,-{1\over2}, -{1\over2})}
for all $\alpha$, corresponding to
gauge invariance in spacetime. A natural solution to the
above constraints which reduces to \spcons\
for $n=0,\pm1$ is:
\eqn\formvir{-L_n=\oint dz\left[
(1-n^2)J^3\gamma^n+{n(n-1)\over2}J^-\gamma^{n+1}
+{n(n+1)\over2}J^+\gamma^{n-1}\right]}
To see that the operators $L_n$
satisfy the Virasoro algebra \virfull\
as well as \kacmd\ it is
convenient to use the gauge invariance
\longphot\ to transform \formvir\ to the equivalent
form:
\eqn\newvir{-L_n=\oint dz
\left[(n+1)J^3\gamma^n-nJ^-\gamma^{n+1}\right]
}
and compute:
\eqn\commvir{[L_n, L_m]=
\oint dw\oint dz
\left[(n+1)J^3\gamma^n-nJ^-\gamma^{n+1}\right](z)
\left[(m+1)J^3\gamma^m-mJ^-\gamma^{m+1}\right](w)
}
There are four terms to evaluate;
the residues of single poles in the OPE are of three different
kinds: $J^3\gamma^{n+m}$, $J^-\gamma^{n+m+1}$
and $\gamma^{n+m-1}\partial\gamma$. The numerical factors
conspire so that the algebra closes. Using the OPE's
\eqn\opes{\eqalign{
J^3(z) \gamma^n(w)=&{n\gamma^n(w)\over z-w}+\cdots\cr
J^-(z) \gamma^n(w)=&{n\gamma^{n-1}(w)\over z-w}+\cdots\cr
J^3(z) J^-(w)=&-{J^-(w)\over z-w}+\cdots\cr
J^3(z) J^3(w)=&-{k/2\over (z-w)^2}+\cdots\cr
}}
one finds that \commvir\
leads to the algebra \virfull\ with
the central charge in spacetime given
in terms of the level of $SL(2,R)$, $k$, \two,
and the charge $p$, \pdef:
\eqn\cspacetime{c_{\rm spacetime}=6kp}
Thus, for fixed $SL(2,R)$ level $k$, as $p$
increases the spacetime central charge
$c_{\rm spacetime}\to\infty$, which is the
semiclassical limit in the spacetime CFT. We will
see later that the string coupling is proportional
to $1/\sqrt p$; thus the theory indeed becomes more
and more weakly coupled as $p\to\infty$. Similarly,
as $k\to\infty$ for fixed $p$, the curvature of $AdS_3$
goes to zero and the gravity approximation to (aspects
of) the full string theory becomes better and better.

Note that the Virasoro algebra acts as holomorphic
reparametrization symmetry on $\gamma$. Indeed, one
can verify using \newvir, \opes\ that:
\eqn\repgam{[L_n,\gamma(z)]=-\gamma^{n+1}(z)}
which implies that one can think of $L_n$ as
\eqn\lng{L_n=-\gamma^{n+1}{\partial\over\partial\gamma}.}

The second line of \kacmd\ is also a straightforward
consequence of \genta, \newvir, \opes. Note that our derivation
of the Virasoro and affine Lie algebras was performed in
the free field limit of \acwak, in which one can ignore the
screening charge $\beta\bar\beta \exp(-2\phi/\alpha_+)$. 
This is accurate at the boundary of 
$AdS_3$, $\phi\to\infty$. One can check that the affine Lie 
and Virasoro generators \genta, \newvir\ do not commute
with the screening charge. This means that there are
corrections to these generators which form a power series in
$\exp(-2\phi/\alpha_+)$. 

In the presence of both $SL(2,R)$ \two\ and
$G$ \glss\ affine Lie algebras on the worldsheet
one can define a second Virasoro algebra in spacetime --
the Sugawara stress tensor of the $\hat G$ generated by
$T^a_n$ \genta. This second Virasoro algebra should be
thought of as a part of the total Virasoro algebra
\formvir. In the three dimensional string theory
the reason for this is that all degrees of freedom
must couple to three dimensional gravity.
In particular, if the spacetime theory is unitary,
the central charges must satisfy the inequality
\eqn\unitrty{c_{\rm spacetime}=6kp\geq{k'p{\rm dim}\;G\over k'p+Q}}
where the right hand side is the Sugawara central charge
for $\hat G$, and $Q$ is the quadratic
Casimir of $G$ in the adjoint representation.
The inequality \unitrty\ becomes trivial
in the weak coupling limit $p\to\infty$, but for
large string coupling $p\simeq 1$ it provides a
constraint on the parameters of the theory. Of course,
the whole discussion of unitarity in the bosonic string is quite
confusing because of the instability which is signaled by the
tachyon.  Below we will apply a similar discussion to stable vacua of
string theory.

The form of the spacetime central charge \cspacetime\
is interesting. In the original work of Brown and Henneaux
\brhen\ this central charge was computed using low energy gravity
and was found to be
\eqn\brhencent{c_{\rm spacetime}={3l\over 2l_p}}
where $l$ is the radius of curvature of $AdS_3$ (see
\metricttheta) and $l_p$ is the three dimensional Planck
length ($l_p\equiv G_3$, the three dimensional Newton constant).
The calculation of \brhen\ is expected
to be reliable in the semiclassical regime
$l\gg l_p$; \brhencent\ should be thought of as the
leading term in an expansion in $l_p/l$.
In our case $l$ is related to
the level of the $SL(2,R)$ affine algebra, $k$ \tb, while
$l_p$ is given in terms of the fundamental string coupling
$g$ and the volume of the compactification
manifold $\NN$ \bosvac, $V_\NN$ (measured in string units), by:
\eqn\lppp{
{1\over l_p}={V_\NN\over g^2 l_s}}
Thus, the formula \cspacetime\ for the central charge
implies in this case that the string coupling is
quantized. More precisely, the three
dimensional Planck scale satisfies:
\eqn\weakcoup{l_s/l_p=4p\sqrt k}
The three dimensional string coupling $g_3^2\sim l_p/l_s\sim 1/(p\sqrt
k)$ is small if either $k$ or $p$ are large. As we will see later, the
higher dimensional string coupling is typically {\it
large}\foot{Essentially because the volume of $\NN$ \bosvac\ typically
grows as $V_\NN\sim k^a$, with $a>1/2$.}  for small $p$, and it
decreases as $p\to\infty$, where perturbative string theory is
reliable.  The fact that the string coupling is fixed in string theory
on $AdS_3$ in a given sector of the theory (\ie\ for given $k, p$)
implies that the dilaton is massive and its potential has a unique
minimum.  

The string coupling behaves like
$g\sim 1/\sqrt p$, which is reminiscent of the
coupling between mesons in large $N$ gauge
theory (where $g\sim 1/\sqrt N$). Perhaps one can think of the
closed strings on $AdS_3$ as ``mesons'' constructed out of
``quarks.''

Another (related) useful analogy is WZW models.
The WZW Lagrangian for a compact group $G$ at level $P$
is proportional to $P$ (at the fixed point where the infinite
conformal and affine Lie symmetries appear). The interactions
between physical states are of order $1/\sqrt P$; $P$ is quantized
due to non-perturbative effects (it must
be a positive integer).
Similarly, in the string theory described above the spacetime
action is proportional to $p$, while interactions are proportional 
to $1/\sqrt p$. The fact that $p$ is quantized
is non-perturbative in the string loop expansion\foot{However,
the discussion
after eq. \pdef\ makes it clear that if $p$ is non-integer
the theory is non-perturbatively inconsistent.}, and presumably
related to the appearance of the infinite symmetry
\virfull, \kacmd\ in spacetime.

Physical states in the theory fall into representations
of the Virasoro algebra \formvir. A large class of such
states is obtained by taking a primary of the worldsheet
conformal algebra on $\NN$ \bosvac, and dressing it
with a conformal primary from the $AdS_3$ sector.
In what follows we will describe this dressing first for
the case of vanishing worldsheet spin and then for non-zero
spin.

Let $W_N$ be a spinless worldsheet operator in the CFT
on $\NN$, with
scaling dimension $\Delta_L=\Delta_R=N$ (which of course need not
be integer). We can form a physical vertex operator
by ``dressing'' $W_N$ by an $AdS_3$ vertex operator
$V_{jm\bar m}$ \primr.
The physical vertex operator
\eqn\physstate{V_{\rm phys}(j,m, \bar m)=W_NV_{jm\bar m}}
must have worldsheet dimension one:
\eqn\loeq{N-{j(j+1)\over k-2}=1}
Stability of the vacuum
requires the solutions of \loeq\ to have real $j$
(see below). Furthermore, the unitarity condition 
\unitaritycon\ shows that only operators with 
$N<(k/4)+1$ can be dressed using \physstate.

To determine the transformation properties
of the spacetime field corresponding to
$V_{\rm phys}$
under the spacetime conformal symmetry,
we need to compute the commutator $[L_n, V_{\rm phys}]$.
A straightforward calculation using the form\foot{One
can also derive this relation by using the representation
\formvir\ and the operator product
$J^\pm(z) V_{j,m,\bar m }(w, \bar
w)=(m\mp j) V_{j, m\pm1, \bar m }(w,\bar w)/(z-w)+ ...$.}
\newvir\ of $L_n$ and the free field realization of
$SL(2,R)$, \acwak, leads to:
\eqn\commfield{[L_n, V_{\rm phys}(j,m, \bar m )]=(nj-m)
V_{\rm phys}(j, m+n, \bar m)}
To understand the meaning of eq. \commfield,
recall the following result from CFT.
Given an operator $A^{(h\bar h )}(\xi,\bar \xi)$ with scaling
dimensions $(h,\bar h)$, we can expand it in modes:
\eqn\modeexp{A^{(h\bar h)}(\xi,\bar \xi)=\sum_{m\bar m} A^{(h\bar h
)}_{m\bar m} \xi^{-m-h}\bar \xi^{-\bar m-\bar h}.}
The precise values of $m$ and $\bar m$ depend,
as usual, on the sector -- the operator insertion
at $\xi=0$. In the identity sector $m+h, \bar m+\bar h \in Z$.
The mode operators $A^{(h\bar h)}_{m\bar m}$ satisfy the following
commutation relations with the Virasoro generators:
\eqn\spavir{[L_n, A^{(h\bar h)}_{m\bar
m}]=\left[n(h-1)-m\right]A^{(h\bar h)}_{n+m, \bar m}.}
Comparing \commfield\ with \spavir\
we see that we should identify the physical vertex
operators $V_{\rm phys}(j,m, \bar m)$
with modes of primary operators in the
spacetime CFT, $A^{(h\bar h)}_{m\bar m}$.
The scaling dimension in spacetime
of the operator $V_{\rm phys}(j,m, \bar m)$ is
$h=\bar h=j+1$:
\eqn\vcorr{V_{\rm phys}(j,m, \bar m )\leftrightarrow
A^{(h\bar h )}_{m\bar m};\;\;\; h=\bar h=j+1}
Note that due to \unitaritycon\ there are 
bounds on the scaling dimensions arising from
single particle states: $0<h < k/2$.
Equation \loeq\ furthermore
relates the spectrum of scaling dimensions to
the structure of the compact CFT on $\NN$.
Tachyons correspond to solutions of \loeq\
with complex $j=-{1\over2}+i\lambda$, and we see \vcorr\ that they
give rise to complex scaling dimensions in
the spacetime CFT. According to \loeq,
worldsheet operators $W_N$ with $N<1-{1\over 4(k-2)}$
in the CFT on $\NN$ correspond to tachyons. 
Since the identity operator
is such an operator that always
exists, bosonic string theory in the background
\bosvac\ is always unstable, just like in flat space.

The operators \physstate\ give rise
to spacetime primaries with spin zero, \ie\
$h=\bar h$, \vcorr. One expects in general to find many
primaries with non-zero spin (in spacetime).
These are obtained by coupling worldsheet operators
with non-zero {\it worldsheet} spin to the $AdS_3$
sector. Consider, for example, a worldsheet primary
$Z_{N,\bar N}$ in the CFT on $\NN$,
with scaling dimensions
$\Delta_L=N$, $\Delta_R=\bar N$. We will assume, without
loss of generality, that $N<\bar N$. We cannot couple
$Z_{N,\bar N}$ directly to $V_{jm\bar m}$ as in \physstate,
since this would violate worldsheet level matching.
In order to consistently couple $Z$ to $AdS_3$, we
need a primary of the worldsheet conformal algebra
on $AdS_3$ that has spin $n=\bar N-N\in Z$, and is thus a descendant
of $V_{jm\bar m}$ \primr\ (under the $SL(2,R)$ current algebra)
at level $n$.

There are many such
descendants; to illustrate the sort of spacetime states
they give rise to, we will study a particular example,
the operator $(\partial_z\gamma)^n V_{jm\bar m}$.
It is not difficult to check that this operator
is a conformal primary with $\Delta_R=-j(j+1)/(k-2)$
and $\Delta_L=\Delta_R+n$. Therefore, as in \physstate,
we can form the physical operator
\eqn\physnew{V^{(n)}_{\rm phys}(j,m,\bar m)=Z_{N,\bar N}
(\partial_z\gamma)^nV_{jm\bar m}}
with
\eqn\levmatch{\bar N-{j(j+1)\over k-2}=1}
Under the right moving
spacetime Virasoro algebra $\bar L_n$, the operator
\physnew\ transforms, as before \physstate,
as a primary with dimension
$\bar h=j+1$.
The addition of $(\partial\gamma)^n$ does
change the transformation of \physnew\
under the left moving Virasoro algebra.
Using eq. \repgam\ we have:
$[L_s,(\partial_z\gamma)^n]= -n(s+1)\gamma^s(\partial_z\gamma)^n$
and, therefore,
\eqn\newcomm{[L_s, V^{(n)}_{\rm phys}(j,m,\bar m)]=
\left[s(j-n)-(m+n)\right]V^{(n)}_{\rm phys}(j,m+s,\bar m)}
Comparing to \spavir\ we see that the left moving spacetime
dimension of our operator is $h=j+1-n$ and the modes $m$ are
shifted by $n$ units. This adds another entry to our
spacetime -- worldsheet correspondence: operators with spin
$n$ on the worldsheet give rise to operators with spin
$n$ in the spacetime CFT.

One might wonder what happens for $j+1-n<0$ when
the left-moving scaling dimension might become
negative. The answer is that the unitarity constraint
\unitaritycon\ does not allow this to happen.
Indeed, $j<(k-2)/2$ implies using
\levmatch\ that $n\le \bar N <(j+3)/2$. Therefore
$j+1-n>(j-1)/2$; it can become negative only for $j<1$.
Furthermore, since $N\ge 0$ and $\bar N\ge N+1\ge 1$, 
\levmatch\ implies that $j\ge 0$. For $j<1$ we have $n<2$,
which leaves only the case $n=1$; therefore, $j+1-n=j\ge 0$.

Note that it is not surprising that we had to use the constraint
\unitaritycon\ to prove that the scaling dimensions in
spacetime cannot become negative, since both have to do
with the unitarity of our string theory in spacetime. 

One can also study the transformation
properties of physical states under the
spacetime affine Lie algebra $\hat G$, \genta.
For example, if the operator $W_N$ in \physstate\ transforms
under the worldsheet affine Lie algebra \glss\
in a representation $R$:
\eqn\mrimkm{K^a(z) W_N(w,\bar w)={t^a(R)\over z-w} W_N(w,\bar w)+...}
where $t^a(R)$ are generators of $G$ in the
representation $R$,
then the physical vertex operator \physstate\
satisfies the commutation relations:
\eqn\comkm{[T^a_n, V_{\rm phys}(j,m, \bar m)]=t^a(R)V_{\rm phys}
(j, m+n, \bar m)}
\ie\ it is in the representation $R$ of the spacetime affine Lie
algebra.

Correlation functions of physical operators
$V_{\rm phys}$
satisfy in this case the Ward identities
of two dimensional CFT. To prove this one uses
\commfield\ and the fact that $L_n|0\rangle=0$
for $n\geq -1$. This last condition can be
understood by thinking of the Virasoro generators
\virgen\ as creating physical states from the
vacuum. Since energies of states are positive definite,
we can think of $L_n$ with $n<-1$ as creation operators,
and of $L_{n>1}$ as annihilation operators. The
latter must therefore annihilate the vacuum.
Note that the identification of observables in
three dimensional string theory with two dimensional 
CFT correlators found here provides a proof (for the 
case of $AdS_3$) of the map between string theory in 
anti-de-Sitter background and boundary CFT proposed 
in \refs{\gkp, \witt}.

We see that string theory on $AdS_3$ has many states
which are obtained by applying the holomorphic
vertex operators
$\oint dz V_{\rm phys}^{(L)}(j=0,m,\bar m=0)$ and their
anti-holomorphic analogs to the vacuum. Examples include the
generators of the spacetime affine Lie \genta\ and
Virasoro \formvir\ algebras. More generally,
since the worldsheet and spacetime chiralities
of operators are tied in this background,
the chiral algebra of the spacetime CFT is
described by states of this form. As we stressed above, these states
are {\it not} standard closed string states.
This new sector in the Hilbert space must be kept even
at large $k, p$, where the theory becomes semiclassical
and the weakly coupled string description is good.
Clearly, these chiral states should also appear
in the discussion of \refs{\gkp, \witt}.

\newsec{Superstrings on $AdS_3$}

Bosonic string theory on $AdS_3$ contains tachyons, which
as we saw means that some of the operators in the spacetime
CFT have complex scaling dimensions, and thus the theory
is ultimately inconsistent. In this section
we generalize the discussion to the spacetime supersymmetric
case, which as we will see gives rise to consistent,
unitary superconformal field theories in spacetime.
We will work in the Neveu-Schwarz-Ramond formalism
\fms.

There are two steps in generalizing
the discussion of section 2 to the supersymmetric
case. The first is introducing worldsheet fermions
and enlarging the worldsheet gauge principle from
$N=0$ to $N=1$ supergravity. This is usually
straightforward, but it does not
solve the tachyon problem.
The second
step involves introducing spacetime fermions
by performing a chiral GSO projection. This leads
to spacetime supersymmetry and a host of new issues,
some of which will be explored below
in the context of
superstring theory on the manifold $\MM $ \one.

\subsec{Fermionic Strings on $AdS_3$}

\nref\kazama{Y. Kazama and H. Suzuki,
\np{321}{1989}{232}.}
Following the logic of section 2, we are interested
in fermionic string propagation in a spacetime
of the form \bosvac.
The $\sigma$ model on $AdS_3\times
\NN$ has $N=1$ superconformal symmetry
with $c=15$ which we gauge
to construct the string vacuum. The worldsheet
SCFT with target space $AdS_3$ will be taken as
before to have an affine $SL(2,R)$ Lie
algebra symmetry at level $k$, generated by
worldsheet currents $J^A$ satisfying the OPE
\two. We will also assume that the SCFT on $\NN$ has in
addition an affine Lie algebra symmetry $\hat G$ corresponding
to some compact Lie group $G$, at level $k'$, with currents $K^a$
and OPE's \glss.  Under the $N=1$ superconformal algebra
on the worldsheet, the currents $J^A$, $K^a$ are upper
components of superfields, whose lower components are
free fermions $\psi^A$ ($A=1,2,3$) and $\chi^a$ ($a=1,
\cdots, {\rm dim}\;G$), respectively (see \eg\ \kazama\
for a detailed discussion of superstring propagation
on group manifolds). The currents $J^A$, $K^a$ can be written
as sums of ``bosonic'' currents $j^A$, $k^a$ whose levels are
$k+2$, and $k'-Q$ (recall that $Q$ is the quadratic Casimir
of $G$ in the adjoint representation, as in \unitrty),
which commute with the free fermions, and contributions
{}from the free fermions which complete the levels to $k$
and $k'$:
\eqn\Jp{\eqalign{
J^A=&j^A-{i\over k}\epsilon^A_{\,\,\,BC}\psi^B\psi^C\cr
K^a=&k^a-{i\over k'} f^a_{\,\,\,bc}\chi^b\chi^c\cr
}}
We are using the convenient but unconventional
normalization of the free fermions,
\eqn\normferm{\eqalign{
\langle\psi^A(z)\psi^B(w)\rangle=&{k\eta^{AB}/2\over z-w},
\qquad A,B=1,2,3\cr
\langle\chi^a(z)\chi^b(w)\rangle=&{k'\delta^{ab}/2\over z-w},
\qquad a,b=1,...,{\rm dim}\, G\cr
}}
As in the bosonic case, the worldsheet currents
\Jp\ lead to spacetime symmetries. To construct
the corresponding charges, recall that in fermionic
string theory, physical states are obtained from
superconformal primaries with dimension $h=1/2$
by taking their upper component
(by applying the $N=1$ supercharge $G_{-1/2}=\oint dw G(w)$),
and integrating the resulting dimension one operator.
In the present case, the gauged
worldsheet supercurrent $G$ is given by
\eqn\gsl{G(z)={2\over k}(\eta_{AB}\psi^Aj^B
-{i\over 3k}\epsilon_{ABC}\psi^A\psi^B\psi^C)
+{2\over k'}(\chi^ak_a
-{i\over 3k'}f_{abc}\chi^a\chi^b\chi^c)+G_{\rm rest}}
where $G_{\rm rest}$ is the contribution to $G$ of the
degrees of freedom that complete \Jp,
\normferm\ to a critical string theory. The relevant dimension
$\half$ superconformal primaries are $\psi^A$ and $\chi^a$.
The corresponding upper components are $J^A$, $K^a$ \Jp,
in terms of which the charges have the same form as in
the bosonic case \spcons, \tkcons, respectively.
In particular, they satisfy the $SL(2,R)\times G$ algebra.

The global symmetry algebra can again be
extended to a semi-direct
product of Virasoro and affine $G$ \virfull, \kacmd.
The $\hat G$ affine Lie algebra generators are:
\eqn\TnGxg{T_n^a=\oint dz \{G_{-1/2},\chi^a\gamma^n(z)\}}
The Virasoro generators $L_n$ are:
\eqn\virnow{\eqalign{-L_n=&\oint dz \left\{G_{-1/2},
(1-n^2)\psi^3\gamma^n+{n(n-1)\over 2}\psi^-\gamma^{n+1}
+{n(n+1)\over 2}\psi^+\gamma^{n-1}(z)\right\}\cr
=&\oint dz \left[(1-n^2)J^3\gamma^n+{n(n-1)\over2}J^-\gamma^{n+1}
+{n(n+1)\over2}J^+\gamma^{n-1}\right]\cr}}
In the second line of \virnow\ we used the fact that all
the terms in which $G_{-{1/2}}$ acts on $\gamma$ cancel. 
Note that this way of writing $L_n$ is the same as \formvir,
which can also be simplified as \newvir. In fact, this result
should have been anticipated because \formvir\ only uses the
presence of $SL(2)$.
We emphasize that in \TnGxg, \virnow\ $G_{-1/2}$ is a
{\it worldsheet} supercharge, while $T^a_n$, $L_n$ are
{\it spacetime} symmetry generators. It is not difficult
to verify by direct calculation that the generators
\TnGxg, \virnow\ satisfy the algebra \virfull, \kacmd,
with the central charges \stlevel, \cspacetime.

Just as in the bosonic case, one can construct
physical states which are primaries of the conformal
algebra \virnow. For simplicity, we describe
the construction for spinless operators
(both on the worldsheet and in spacetime).
These are obtained by taking a primary of the
$N=1$ worldsheet superconformal algebra on $\NN$, $W_N$,
with scaling dimension $\Delta_L=\Delta_R=N$, and
dressing it with a superconformal primary on $AdS_3$.
The corresponding vertex operator in the $-1$ picture is:
\eqn\physferm{V_{\rm phys}(j,m, \bar m )=e^{-\phi -\bar \phi
}W_NV_{jm\bar m}}
where $\phi$ and $\bar \phi$ are the bosonized super-reparametrization
ghosts\foot{In section 2 we denoted by $\phi$ the radial
direction in $AdS_3$ \uycoo. It should be clear from
the context which $\phi$ we mean everywhere below.}
\fms. The commutation relations of the operators 
\physferm\ with the Virasoro generators  $[L_n, V_{\rm phys}]$ 
are similar to the bosonic case \commfield. The resulting
scaling dimensions are:
\eqn\scdim{h=\bar h=j+1;\;\;\; N-{j(j+1)\over k}={1\over2}}
In particular, states with $N<{1\over2}-{1\over 4k}$ (tachyons)
give rise to complex scaling dimensions. The lowest such state
is the identity $W_N=1$. Its presence in the
spectrum implies that the fermionic string on $AdS_3
\times \NN$ is an unstable vacuum of string theory, just like
the bosonic theory of section 2.
However, in this case there is a well known solution to
the problem. One can eliminate the tachyons from the spectrum
by performing a chiral GSO projection. We will next describe
this projection for the particular case of $AdS_3$.
Rather than being very general,
we will do that in the context of an example: superstring theory
on $AdS_3\times S^3\times T^4$ \one.

\subsec{Superstrings on $\MM=AdS_3\times S^3\times T^4$}

In addition to $AdS_3$, the manifold $\MM$ includes now
a three-sphere, or equivalently the $SU(2)$
group manifold. The worldsheet theory is the
$N=1$ superconformal
WZW model on $S^3$. We use the notation of \Jp, \normferm.
The $AdS_3$ fermions and currents are denoted
by $(\psi^A, J^A)$, while those corresponding to $SU(2)$
are $(\chi^a, K^a)$. The levels of $SL(2)$ and $SU(2)$
current algebras are $k$ and $k'$, respectively.

The total central charge of the $AdS_3\times
S^3$ part of the worldsheet SCFT is:
\eqn\cc{c={3(k+2)\over k}+{3\over 2}+{3(k'-2)\over k'}+{3\over 2}}
The first and third terms on the r.h.s.
of \cc\ are the contributions of the
bosonic $\sigma$ models on $AdS_3$ and $S^3$;
the second and fourth are due to
worldsheet fermions. Criticality of the fermionic
string in the background \one\ implies that
$c=9$, which leads to a relation between the
levels of the curent algebras:
\eqn\kprimek{k^\prime=k}
The $T^4$ in $\MM$ corresponds to an
$N=1$, $U(1)^4$ SCFT; four canonically
normalized (compact) free scalar fields
$Y^i$ and fermions $\lambda^i$, $i=1,2,3,4$.
The energy-momentum tensor $T(z)$ and supercurrent
$G(z)$ of this system are:
\eqn\TG{
\eqalign{
T(z)=&{1\over k}(j^Aj_A-\psi^A\partial\psi_A)
              +{1\over k}(k^ak_a-\chi^a\partial\chi_a)
+{1\over2}(\partial Y^i \partial Y_i-
\lambda^i\partial\lambda_i)\cr
G(z)=&
{2\over k}\left(\psi^Aj_A-{i\over 3k}\epsilon_{ABC}\psi^A\psi^B\psi^C\right)
+{2\over k}\left(\chi^ak_a-{i\over 3k}\epsilon_{abc}\chi^a\chi^b\chi^c\right)
+\lambda^i\partial Y_i\cr
}}
So far our treatment of string theory in the background
\one\ was a special case of the discussion of the previous
subsection and, in particular, the resulting spacetime theory
is tachyonic.
We would like next to perform a chiral GSO projection
and remove the tachyons, in the process making the
vacuum supersymmetric. We expect
to be able to perform different projections, corresponding
to different boundary conditions for the spacetime supercharges.
We start with the construction of the vacuum corresponding
to the NS sector of the spacetime SCFT, and then turn to the
Ramond vacuum.

\medskip
\noindent
{\it 1) The Neveu-Schwarz Sector of the Spacetime Theory}

\noindent
A well known sufficient condition
for spacetime supersymmetry
is the enhancement of the $N=1$
superconformal symmetry of the
worldsheet theory to $N=2$ superconformal.
This requires in particular the existence
of a conserved $U(1)_R$ current in the
worldsheet theory, under which the supercurrent
$G$ splits into two parts, $G=G^++G^-$ with
charges $\pm 1$. It will turn out that this
standard route is {\it not} the way to proceed
here\foot{In Appendix B we describe some features
of the theory obtained by utilizing an $N=2$ superconformal
symmetry on the worldsheet, and its relation
to the spacetime SCFT studied in this section.}.
We will next construct the spacetime supercharges
directly, and study the resulting superalgebra.

It is convenient to start by pairing the ten
free worldsheet fermions (\ie\ choosing a complex
structure) and bosonizing them into five
canonically normalized scalar fields,
$H_I$, $I=1,...,5$, which satisfy
$\langle H_I(z)H_J(w)\rangle=-\delta_{IJ}\log(z-w)$:
\eqn\HHHHH{\partial H_1={2\over k}\psi^1\psi^2, \quad
\partial H_2={2\over k}\chi^1\chi^2, \quad
i\partial H_3={2\over k} \psi^3\chi^3, \quad
\partial H_4=\lambda^1\lambda^2, \quad
\partial H_5=\lambda^3\lambda^4}
All the fields except for $H_3$ satisfy
the standard reality condition $H_I^\dagger=H_I$.
Because of the negative norm of $\psi^3$, $H_3$
instead satisfies $H_3^\dagger=-H_3$.
Note that the standard ``fermion number'' current
$J=i\sum_I \partial H_I$ is not suitable for
extending the $N=1$ superconformal algebra \TG\
to $N=2$, since the OPE of $J(z)$ with $G(w)$ contains
a double pole from the second and fourth terms in $G$ (see
Appendix B for a discussion of the $N=2$ superconformal
structure on the worldsheet).

Ignoring this complication and proceeding,
following the most naive version of \fms,
we attempt to construct supercharges of
the form
\eqn\supchh{Q=\oint dz e^{-{\phi\over2}}S(z); \;\;\;
S(z)=e^{{i\over2}\sum_I\epsilon_IH_I}}
where $\epsilon_I=\pm 1$. In flat space all
$32$ supercharges \supchh\ are BRST invariant
and due to the requirement of mutual locality
between different supercharges, which is necesssary
to have a sensible worldsheet theory, one keeps only the sixteen
supercharges that satisfy
\eqn\flatc{\prod_{I=1}^5\epsilon_I=1}
Since we did not use an
$N=2$ superconformal algebra on the worldsheet to construct 
the supercharges, in our case BRST invariance
of \supchh\ is not guaranteed. Indeed, due to the
second and fourth terms in $G$ \TG\ one finds that
only the supercharges that satisfy in addition to
\flatc,
\eqn\flatca{\prod_{I=1}^3\epsilon_I=1} are physical.
Thus, this system has eight spacetime supercharges (from
each worldsheet chirality). The supercharges
that survive \flatc, \flatca\ can be labeled
by three signs, say $\epsilon_1$, $\epsilon_2$
and $\epsilon_4(=\epsilon_5)$. The meaning of these
signs is revealed by looking at the transformation
of the supercharges under the bosonic symmetries
of the vacuum, $SL(2,R)\times SU(2)\times SO(4)$,
with the last factor rotating the four fermions
$\lambda^i$. The supercharges are in the $({\bf
\half }, {\bf \half }, {\bf\half}, {\bf 0} )$ of this
symmetry. Thinking
about the $SL(2,R)$ charges as the global part of
a spacetime Virasoro algebra, we see that we have
four pairs of supercharges
$Q_{\pm {1\over2}}^{\epsilon_2,\epsilon_4}$ in
the $({\bf \half}, {\bf\half}, {\bf 0})$ of $SU(2)\times SO(4)$.

Using the technology of \fms\ one finds that
the anticommutators of $Q_r^{\epsilon_2,\epsilon_4}$
($r=\pm 1/2$, $\epsilon_2, \epsilon_4=\pm 1$) close
on the $SL(2,R)\times SU(2)$ charges \spcons, \tkcons.
The superalgebra formed by these objects is the global
$N=4$ superconformal algebra in the NS sector:
\eqn\nssup{
\eqalign{
\{Q_r^i,\bar Q_s^j\}&=2\delta^{ij}L_{r+s}
-2(r-s)\sigma_{ij}^aT^a_{r+s} \cr
\{Q_r^i,Q_s^j\}&=0=\{\bar Q_r^i,\bar Q_s^j\},
\qquad i,j=1,2,\quad r,s=\pm 1/2 \cr
[T_0^a,Q^i_r]&=-{1\over 2}\sigma_{ij}^a Q_r^j, \qquad
[T_0^a,\bar Q_r^i]={1\over 2}\sigma_{ij}^{a*} \bar Q_r^j \cr
[L_n,Q_r^i]&=({n\over 2}-r)Q_{n+r}^i, \qquad
[L_n,\bar Q_r^i]=({n\over 2}-r)\bar Q_{n+r}^i, \qquad n=0,\pm 1
}}
where we have formed out of our supercharges
\supchh\ two $SU(2)$ doublets (for given
$SL(2,R)$ weight $r$). $Q_r^i$ in \nssup\
corresponds in the language of \supchh\ to $\epsilon_1=2r$,
$\{\epsilon_2=\pm 1\}\leftrightarrow\{i=1,2\}$,
$\epsilon_4=1$, and $\bar Q_r^i$ is the same but
with $\epsilon_4=-1$; $\sigma^a$ are Pauli matrices.
The commutation relations of the supercharges with
$L_n$, $n=0,\pm1$, and $T^a_0$, $a=1,2,3$, are determined
by recalling that the supercharges have scaling dimension
$h=3/2$ in spacetime and \spavir, and that they transform
in the ${\bf 2}$
of the spacetime $SU(2)$ symmetry and \comkm.

As we saw in the previous subsection, the bosonic
part of the superalgebra \nssup, namely
$SL(2)\times SU(2)$, is extended
to a semi-direct product of a $c=6kp$ Virasoro algebra
and an affine $SU(2)$ at level $kp$. This clearly means that
the $N=4$ supercharges $Q_r$ which we have constructed
only for $r=\pm1/2$, actually exist with an arbitrary
$r\in Z+1/2$. One way of finding them is to act with
$L_n, T^a_n$ \TnGxg, \virnow\ on $Q_{\pm 1/2}$.
The resulting structure is the full $N=4$ superconformal
algebra in spacetime.

Note also that, as in the bosonic case, there is
a correlation between chirality on the worldsheet
and in spacetime. The spacetime dynamics is that
of a $(4,4)$ superconformal field theory, with the
right moving chiral algebra in spacetime arising
{}from the right movers on the worldsheet via formulae
like \TnGxg, \virnow, \supchh, and similarly for the
left movers.

There are now two kinds of physical states.
Bosons are described by vertex operators
of the form \physferm. The fermion vertices
are straightforward generalizations of those
described in \fms. They are related
to the bosons by supersymmetry \supchh\ --
\nssup. Additional bosonic states appear from the
worldsheet RR sector.

We will only comment briefly on the spectrum of
physical states, leaving a more detailed analysis
to future work (see also section {\it 4.3}). Consider the vertex
operators \physferm. The string theory has eight towers
of oscillator states coming from all three sectors
in \one: the $AdS_3$, $SU(2)$ and $T^4$ parts
of the worldsheet SCFT.
Roughly speaking, four of the eight towers can be thought
of as describing descendants of the $N=4$ superconformal
algebra in spacetime \nssup. The four remaining
towers correspond to descendants of the $U(1)^4$
affine Lie algebra generated by the operators
\eqn\spn{\alpha_m^i=\oint dz e^{-\phi}\lambda^i\gamma^m}
which satisfy the commutation relations \kacmd\
\eqn\commrelt{[\alpha_n^i,\alpha_m^j]=pn
\delta^{ij}\delta_{n+m,0}}
Examples of low lying physical states are the scalar
fields $B^{i\bar i}$ obtained by Kaluza-Klein reduction
of the metric and $B$ field from ten down to six dimensions
on $T^4$.
The $l=2j$ partial
wave of $B^{i\bar i}$ on the sphere
transforms in the $({\bf j}, {\bf j})$
representation of the $SU(2)_L\times SU(2)_R$
isometry of $S^3$ and is
described by the vertex operator
\eqn\bij{B^{i\bar i}(j;m,m',\bar m , \bar m')=e^{-\phi-\bar \phi
}\lambda^i\bar \lambda^{\bar i} V_{jm \bar m}
V'_{j m' \bar m'}}
where $V'_{j'm' \bar m'}$ is the vertex operator of the
$SU(2)$ WZW model with isospin
$j'$, $j'_3=m'$, $\bar j'_3= \bar m'$, and we have set
$j=j'$ in  \bij\ to satisfy \scdim. The scaling
dimension of the operator $B^{i\bar i} (j)$ is \scdim,
$h=\bar h=1+j=1+l/2$. $j$ takes the values\foot{Note that
this is consistent with \unitaritycon\ since we have to replace
$k\to k+2$ there to account for the difference between the
full and bosonic $SL(2)$ level.}
$j=l/2$, $l=0,1,2,\cdots, k-2$. 
Applying the superconformal
and $U(1)^4$ generators to \bij\ generates the spectrum
of the theory.

The states obtained by acting on the massless vertex
operators \bij\ with the spectrum generating operators
\spn\ can be alternatively described by replacing $\lambda^i$
in \bij\ by excited state vertex operators in the $(4,4)$ supersymmetric
worldsheet theory on $T^4$. This confirms that there are
four towers of such (single particle) states.

Spacetime fermion vertices are obtained as usual \fms\
by acting on the boson vertex operators, \eg\ \bij,
with the spacetime supercharges \supchh\ -- \flatca.
This gives rise to vertex operators in the $-3/2$ picture,
which can be brought to the standard $-1/2$ picture
by applying the picture changing operator of \fms.

Some of the resulting spacetime fermions correspond to chiral
operators in the spacetime SCFT in the sense that their $SU(2)$
quantum number coincides with their dimension. For example, the
superpartners of the partial waves of the six dimensional massless
scalar $B^{i\bar i}$ \bij\ are complex fermions
$F^{a\bar i}(j+{1\over2}; r,r',\bar m, \bar m')$
carrying a spinor index $a$ under $SO(4)$ and
spin $j+{1\over2}$ under $SU(2)$. $r$ and $r'$ are the eigenvalues of
$-L_0$ and $T^3_0$. 
Since $B^{i\bar i} (j)\sim
Q_{-1/2}F^{a \bar i}(j+{1\over2})$, the scaling dimension of the
fermions $F$ in the spacetime SCFT is $h_F=j+{1\over2}$. Thus, these
operators have the property that their scaling dimension and $SU(2)$
spin coincide -- they are chiral in spacetime. This gives rise to $k-1$
(complex) chiral operators with $SU(2)$ spins $j=n/2$, $n=1,2,\cdots,
k-1$.  Of course, we can also apply the supercharges with the
opposite spacetime chirality and construct bosons
$B^{a\bar a}$ which are chiral under both the left moving and the right
moving spacetime superconformal algebras.

\nref\malstr{J. Maldacena and A. Strominger, hep-th/9804085.}%

{}From the general representation theory of
$N=4$ SCFT, in a unitary theory with $c=6kp$ we expect to find chiral
operators in small representations
with $SU(2)$ spins $j\leq kp/2$.
The states with $k/2\le j\le kp/2$ correspond in
string theory to multiparticle states.
For finite $p$ the spectrum of multiparticle
chiral states is truncated at $j=kp/2$.
This is reminiscent of a similar
phenomenon in WZW theory. Classically, the WZW Lagrangian
describes an infinite number of primaries of $\hat G$,
while quantum effects restrict the possible representations,
\eg\ in the case of $G=SU(2)$ to $j\le k/2$. Like here, the
simplest way to see this restriction is to impose unitarity
of the quantum theory. The truncation of the
spectrum of multi-particle chiral operators in string
theory on $\MM$ has been recently discussed in \malstr.

\medskip
\noindent
{\it 2) The Ramond Sector of the Spacetime Theory}

\noindent
Having understood the string vacuum corresponding to
the NS sector of the spacetime SCFT we next turn to the
Ramond vacuum.
In this vacuum we expect to find integer modded spacetime
supercharges $Q^i_n$, $\bar Q^j_m$ with $i,j=1,2$ and
$n,m\in Z$, satisfying the algebra \nssup\ (and a similar
structure from the other spacetime chirality).

Since the Euclidean $AdS_3$ space \defeq\ is simply connected,
it has only one spin structure.
Spinors do not change sign when transported around any point, say
$\gamma =0$, with any $\phi $ (in the notation
\uycoo\ -- \inversecha). The change of variables \changeva\
leads to a change of sign when spinors are transported around
$\gamma=0$, \ie\ under $\theta \rightarrow \theta+2\pi$.  In terms of
a SCFT on the boundary, string theory on
$AdS_3$ thus corresponds to the NS sector.  If
we want to describe the R sector we need to introduce two spin fields
in the boundary field theory.  We can put them at $\gamma=0$ and
$\gamma=\infty$, \ie\ at $\tau=\pm \infty$.  These two points on the
boundary can be connected by a line through the bulk.  
The line going through the bulk is the
(analytic continuation to Euclidean space of the) worldline of an
$M=J=0$ black hole.
With this line
omitted from the space, the latter is no longer simply connected and
there can be a non-trivial spin structure. In the R sector, fermions
are antiperiodic under $\gamma\to e^{2\pi i}\gamma$.  

\nref\schwsei{A. Schwimmer and N. Seiberg, \pl{184}{1987}{191}.}
To construct the Ramond sector of the spacetime SCFT using worldsheet
methods, we use the isomorphism (a.k.a. spectral flow) \schwsei\ of
the NS and R $N=4$ superalgebras \nssup.  Given generators that
satisfy the NS algebra \nssup, we can define a one parameter set of
algebras labeled by a variable $\eta$, which for $\eta=1$ (say) goes
over to the NS algebra, and for $\eta=0$ to the Ramond one. Some of
the generators in \nssup\ have $\eta$ dependent modes. If we denote
the generators of the algebras by $L_n(\eta)$, \etc., the generators of
the algebra at $\eta$ are given in terms of the $\eta=0$ (Ramond)
generators by \schwsei:
\eqn\spectflow{
\eqalign{
&T^3_n(\eta)=T^3_n(0)-{\eta kp\over2}\delta_{n,0};\;\;\;
T^\pm_{n\pm \eta}(\eta)=T^\pm_n(0)\cr
&Q^1_{n+{\eta\over2}}(\eta)=Q^1_n(0);\;\;\;
Q^2_{n-{\eta\over2}}(\eta)=Q^2_n(0)\cr
&L_n(\eta)=L_n(0)-\eta T^3_n(0)+\eta^2{kp\over 4}\delta_{n,0}\cr
}}
One can verify that the generators \spectflow\ indeed
satisfy the $N=4$ superconformal algebra \nssup.

Therefore, the operators that we have constructed
before, \TnGxg, \virnow,
\supchh\ -- \flatca, that were interpreted
as generating an NS $N=4$ superalgebra \nssup,
can be reinterpreted as generators of the Ramond
superalgebra using the dictionary \spectflow\ with
$\eta=1$. Denoting the charges which generate the
Ramond superalgebra by $\tilde L_n$, $\tilde T^a_n$,
$\tilde Q_n$, \etc., we have for example:
\eqn\mapops{\eqalign{
\tilde T^3_n=& T^3_n +{kp\over2} \delta_{n,0} \cr
\tilde T^+_n=& T^+_{n+1};\;\;\; \tilde T^-_n=T^-_{n-1}\cr
\tilde L_n=& L_n+T^3_n +{kp\over4}\delta_{n,0}\cr
}}
The shifts in $T^3_0$ and $L_0$ (first and third
lines of \mapops) are due to the fact that the Hilbert
space also transforms non-trivially under spectral flow.
For example, the NS vacuum, which is annihilated by $L_0$, $T^3_0$,
is mapped to a Ramond ground state with $\tilde L_0=kp/4(=c/24)$
and the largest possible $SU(2)$ charge, $\tilde T^3_0=kp/2$.

The excitations of the Ramond ground states are given,
as before, by vertex operators such as \bij.
Using the redefinition of the Virasoro generators
\mapops, as well as the commutation relations
\commfield, \comkm, one finds that physical operators
such as $B^{i\bar i}(j;m,m', \bar m,\bar m' )$ \bij\ satisfy the
commutation relations
\eqn\newvircom{[\tilde L_n, B^{i\bar i}(j;m,m', \bar m,\bar
m')]=\left(nj-(m-m')\right) B^{i\bar i} (j;n+m,m', \bar m,\bar m')}
Comparing to \spavir\ we see
that the fields still have the same
scaling dimensions $h=\bar h=j+1$, but their modes
are shifted by $m'$ (the $SU(2)$ charge $T^3_0$).
The chiral operators $B^{a\bar a} (j+{1\over2}; r,r',\bar r,\bar r')$
with $j\in Z+{1\over2}$ acquire
zero modes, corresponding  to $r=r'$ (and thus
$\tilde L_0-kp/4=0$) and/or $\bar r=\bar r'$.
Therefore, the Ramond vacuum is highly degenerate.

\newsec{Applications of String Theory on $\MM$}

In this section we explain the relation of
string theory on $AdS_3$ discussed here to
other problems of recent interest.

\subsec{Relation to the Theory of the NS Fivebrane}

\nref\chs{C. Callan, J. Harvey and A. Strominger,
\np{359}{1991}{611}; \np{367}{1991}{60}.}%
\nref\dabhar{A. Dabholkar, G. Gibbons, J. Harvey and
F. Ruiz-Ruiz, \np{340}{1990}{33}.}%
\nref\tsey{A. Tseytlin, hep-th/9601177, \mpl{A11}{1996}{689}.}%
Callan, Harvey and Strominger (CHS) \chs\
found the classical supergravity fields around $k$ $NS5$-branes.
One can wrap the fivebranes on a four-torus of arbitrary
volume\foot{We will mostly ignore other moduli of the torus,
and possible background RR fields. The full moduli space
will be discussed in section {\it 4.3}.} $vl_s^4$, parametrized by
the coordinates $x^i$, $i=1,2,3,4$; the fivebranes then
become $k$ strings
whose worldsheet is in the $(\gamma, \bar \gamma)$ plane.
One can extend the CHS solution by adding $p$ fundamental strings
(which are ``smeared over the four-torus'') parallel to the
fivebranes. The supergravity solution corresponding to $p$ fundamental
strings (and $k=0$) was found in \dabhar; the solution with general
$p$ and $k$ was found in \tsey.
The dilaton, NS $B_{\mu\nu}$ field and metric corresponding
to this collection of fivebranes and strings are:
\eqn\dfdoss{
\eqalign{
& {1 \over g_{eff}^2(r)} = e^{-2 \Phi} = {1 \over g^2} f_5^{-1} f_1
\cr
& H= 2 ik  \epsilon_3 + {2i g^2 p \over v} f_5f_1^{-1} *_6
\epsilon_3 \cr
&{ds}^2 = f_1^{-1} d\gamma d\bar \gamma + dx_idx^i + f_5 ( dr^2 + r^2
d\Omega_3^2) \cr}}
The two contributions to the Kalb-Ramond field in the second line
of \dfdoss\ correspond to the $k$ fivebranes and $p$ strings, respectively.
$g$ is the (arbitrary)
string coupling at infinity, $*_6$ is the Hodge dual in the six
dimensions $\gamma, \bar \gamma, r, \Omega_3$, and
\eqn\foffd{\eqalign{
&f_1= 1+{g^2 l_s^2p\over vr^2}\cr
&f_5= 1+{l_s^2k\over r^2}.\cr}}
The first line of eq. \foffd\ takes into account the smearing
of the fundamental string charge over the four-torus. It is valid
(for a torus which is roughly square) for $r\gg v^{1\over4}l_s$.

Note that in the classical limit, $g \rightarrow 0$, the solution goes
over to that of CHS \chs.  In this limit the $k$ $NS5$-branes affect the
background fields because they are heavy (their tension scales like $1
\over g^2$), while the effect of the fundamental strings (whose tension
is of order one) goes to zero. If we want to retain the effect of the
fundamental strings in the classical limit, we have to take
$p \rightarrow \infty$ with $g^2p$ fixed.  Intuitively, the mass of
the $p$ strings then 
scales like $1 \over g^2$ and, therefore, they can affect
the background.

\nref\tseycv{M. Cvetic and A. Tseytlin, hep-th/9510097,
\pl{366}{1996}{95}; hep-th/9512031, \prd{53}{1996}{5619}.}%
We can now study the near horizon geometry of \dfdoss, which corresponds
to distances $r$ which satisfy:
\eqn\nearh{{g^2l_s^2p\over vr^2}\gg1;\;\;\;{l_s^2k\over r^2}\gg1}
Since the validity of \foffd\ requires
$r\gg v^{1\over4}l_s$, we conclude that to study the near horizon
region \nearh\ in a weakly coupled theory we must have $p\gg v^{3\over2}$,
$k\gg v^{1\over2}$. In the limit \nearh\ the configuration
\dfdoss\ turns to:
\eqn\dfdossn{
\eqalign{
& {1 \over g_0^2(r)} = e^{-2 \Phi_0} = {p \over v k}
\cr
& H_0= 2 i k ( \epsilon_3 +  *_6 \epsilon_3 ) \cr
&{ds}_0^2 = k {r^2 \over l_s^2} d\gamma d\bar \gamma + kl_s^2 ({1
\over r^2} dr^2 +d\Omega_3^2)+  dx_idx^i  \cr}}
where we have rescaled $\gamma$ and $\bar\gamma$.  We would like to
make a few comments about this solution:
\item{(a)}  Unlike the solution \dfdoss, here the string coupling is a
constant independent of $r$.  Its value is independent of the 
coupling at infinity, $g$. Thus the dilaton is a fixed scalar.
\item{(b)}  The number of strings
$p$ enters only in the string coupling constant.
Furthermore, the string coupling depends on $p$
in exactly the way that was needed above in order for the fundamental
strings to affect the background, \ie\ $g^2 \sim {1 \over p}$.
The six dimensional string coupling
\eqn\sixdcou{{1 \over g_6^2} = {v \over g_0^2} = {p \over k}}
is independent of $v$.
\item{(c)}  The moduli space of classical solutions such as \dfdoss\ is
subject to some stringy identifications.  For example, the action of
T-duality on the four-torus includes the transformation $v
\rightarrow 1/v$. Therefore, we can limit ourselves to $v \ge 1$.
\item{(d)}  The configuration \dfdossn\ is precisely the one we
studied\foot{The role of $SL(2)\times SU(2)$ in describing the 
near-horizon geometry of \dfdoss\ was pointed out in \refs{\tsey, \tseycv}.}
in section 3.  Here we see how it is embeded in a CHS-like
solution \dfdoss\ which is asymptotically flat.  This provides further
evidence for the interpretation of $p$ defined in \pdef\ as the
number of fundamental strings in the background.
\item{(e)}  It is important to identify the range of validity of the
analysis in section 3.  The worldsheet
theory is weakly coupled for $k \gg 1$.  However, most of our
analysis in section 3 treats the CFT exactly and, therefore, does not
depend on this condition.  For the strings to be weakly coupled we
need $g_0^2(r) = {vk\over p} \ll 1$.

\subsec{Relation to the D1/D5 System}

\nref\calmal{C. Callan and J. Maldacena, hep-th/9602043, 
\np{472}{1996}{591}.}%
\nref\hms{G. Horowitz, J. Maldacena and A. Strominger,
\pl{383}{1996}{151}, hep-th/9603109.}%
The field configuration corresponding to $k$ $NS5$-branes
and $p$ fundamental strings \dfdoss\ is mapped under S-duality
into that describing $k$ $D5$-branes and $p$ $D$-strings \calmal:
\eqn\dfdossd{\eqalign{
& {1 \over \hat g_{eff}^2(r)} = g_{eff}^2(r) = e^{-2\hat
\Phi} = g^2 f_1^{-1} f_5 \cr
&\hat H=H= 2i k  \epsilon_3 + {2i p g^2\over v} f_5f_1^{-1} *_6
\epsilon_3 \cr
&\widehat{ds}^2 = e^{-\Phi}{ds}^2={1\over g} f_1^{-\half}f_5^{-\half}
d\gamma d\bar \gamma + {1\over g}  f_1^\half f_5^{-\half}  dx_idx^i +
{1\over g}  f_1^\half f_5^\half (dr^2 + r^2 d\Omega_3^2).\cr}}
Its near horizon limit  is
\eqn\dfdossdn{\eqalign{
& {1 \over \hat g_0^2(r)} = e^{-2\hat \Phi_0} = {kv \over p} \cr
&\hat H_0 = 2i k  (\epsilon_3 + *_6 \epsilon_3)  \cr
&\widehat{ds}_0^2 = {r^2 \over l_s^2} \sqrt{v \over kp}d\gamma d\bar
\gamma  +   \sqrt{p\over kv} dx_idx^i +
\sqrt{kp \over v} l_s^2 ({ 1 \over r^2} dr^2 + d\Omega_3^2).\cr}}
A few comments are in order:
\item{(a)}
After rescaling all the coordinates in \dfdossd\ by $g^\half$ we find
the D1/D5 solution of \refs{\calmal, \hms}
with the identifications $Q_5=k$ and $Q_1=p$.
\item{(b)}
The direct relation between \dfdossn\ and \dfdossdn\ was obtained in
\malstr.  Our minor addition to their calculation is the observation
that S-duality commutes with taking the near horizon limit of
\dfdoss.
\item{(c)}
In the $D$-brane picture the volume of the four-torus and the six and
ten dimensional string couplings are
\eqn\fourtorvol{\eqalign{
&\hat v = {p \over k} \cr
&{1 \over \hat g_0^2} = {kv \over p} \cr
&{1 \over \hat g_6^2} ={\hat v \over \hat g_0^2} = v .\cr}}
The free continuous parameter of the $NS$ problem, $v$, is now
interpreted as the six dimensional coupling constant, while the
volume of the four torus $\hat v$ is fixed in terms of $p$ and $k$.
\item{(d)}
The parameter space of the problem is subject to discrete identifications.
For example, T-duality includes the transformation $p \leftrightarrow k$.
Therefore, we can limit ourselves to $p \ge k$.
\item{(e)}
String loop corrections are small in the $D$-brane picture when
$\hat g_0^2 = {p \over kv} \ll 1$.  The worldsheet theory is weakly
coupled (the low energy supergravity is a good approximation) when
also ${pk \over v} \gg 1$.  Clearly, there is no situation where both
the $NS$ and $D$ descriptions are simultaneously weakly coupled.

\nref\emil{E. Martinec, hep-th/9804111.}%
\nref\bbg{K. Behrndt, I. Brunner and I. Gaida, hep-th/9804159.}%
\nref\dkss{S. Deger, A. Kaya, E. Sezgin and P. Sundell,
hep-th/9804166.}%
\nref\vaf{C. Vafa, hep-th/9804172.}%
\nref\bbkho{M. Banados, K. Bautier, O. Coussaert, M.
Henneaux and M. Ortiz, hep-th/9805165.}%
\nref\jdb{J. de Boer, hep-th/9806104.}%

Maldacena \malda\ proposed that string theory in the near-horizon
background \dfdossd\ describes in spacetime the CFT obtained
by studying the extreme IR dynamics of $p$ $D$-strings and
$k$ $D5$-branes (see also \refs{\malstr, \emil-\jdb} for more
recent work on this correspondence).
This system corresponds to a $(4,4)$ SCFT
with central charge\foot{Actually, $c=6(kp+1)$, but a
$c=6$ part of the theory is free and decoupled; it plays no role
in the subsequent discussion and thus will be ignored.}
$c=6kp$. It has two decoupled
sectors which are sometimes referred to as the CFT's of
the Coulomb and Higgs branches. The former is obtained
by quantizing the motion of the $D$-strings away from the
$D5$-branes in the directions transverse to both; the latter
corresponds to the $D$-strings being absorbed by the $D5$-branes,
becoming $p$ small instantons in $U(k)$ and growing
to finite size instantons.

The theory of the Higgs branch (which we will refer
to as the $D1/D5$ SCFT) is of interest for
applications, such as the calculation of the Bekenstein-Hawking
entropy of five dimensional black holes \strvafa, as well as
three dimensional ones \strom, the matrix description
of certain non-critical string theories and the $(2,0)$ theory
\ref\matrixm{E. Witten,
hep-th/9707093; O. Aharony, M. Berkooz, S. Kachru,
N. Seiberg, and
E. Silverstein,
hep-th/9707079.},
\etc, and is the one that, according to \malda, is described
by string theory on \dfdossd. 

As we saw here, there are actually two weakly coupled
descriptions of the $D1/D5$ SCFT, each useful in a different
region in parameter space. The theory depends on the discrete
parameters $p$, $k$, and on continuous moduli like $v$.
This parameter space is subject to discrete identifications
such as $v\rightarrow {1 \over v}$, $p \leftrightarrow k$, \etc.
These identifications can be used to restrict to $p\ge k$, $v\ge 1$.
For some range of parameters ($p \gg vk$) the $NS$ description  is weakly
coupled and useful.  For $p \ll vk$ the $D$ description is weakly
coupled but it requires an understanding of string
theory in RR background fields.  If also $pk
\gg v$, one can use supergravity to understand many
aspects of the physics.  Most of the existing work on this system is
in this regime. For generic $p,k,v$ the theory
is strongly coupled in all of the above descriptions.

\subsec{Comparison of String Theory on $\MM$
and the $\sigma$ Model on $T^{4kp}/S_{kp}$}

\nref\cumrun{C. Vafa, hep-th/9511088, \np{463}{1996}{415};
hep-th/9512078, \np{463}{1996}{435}.}%
\nref\rDVV{R. Dijkgraaf, E. Verlinde and H. Verlinde,
hep-th/9603126,
\np{486}{1997}{77};
hep-th/9604055, \np{486}{1997}{89}; hep-th/9704018, \np{506}{1997}{121}.}%
\nref\wadia{S. Hassan and S. Wadia, hep-th/9712213.}%
In the previous section we described
some of the features of the spacetime SCFT corresponding
to fundamental string theory on $\MM$ \one; as explained
above, it is the same SCFT as the $D1/D5$ system. This
SCFT is expected to be equivalent to a $(4,4)$ $\sigma$ 
model on the target space
\eqn\targets{\PP=(\tilde T^4)^{kp}/S_{kp}}
at some value of the moduli which resolve
the orbifold singularity \refs{\cumrun, \strvafa, \rDVV}
(see also \wadia). $\tilde T^4$ must be distinguished
{}from the four-torus $T^4$ on which the fundamental
strings propagate. In this subsection we will
comment on the relation between the spacetime SCFT corresponding
to string theory on $\MM$ and
the $\sigma$ model on $\PP$ \targets;
a precise study of the relation is left for future work.
Since the NS and R sectors of any $(4,4)$ SCFT are equivalent
by spectral flow \spectflow,
we will restrict our comments to the
NS sector of the spacetime theory.

First, note that the two theories have the same chiral
algebra. In addition to the $(4,4)$ superconformal algebra
they also share a $U(1)^4$ affine Lie superalgebra
(actually two copies of $U(1)^4$ from the two chiralities).
In string theory on $\MM$ this symmetry is generated by
the vertex operators $\oint e^{-\phi} \lambda^i \gamma^m$
\spn. In the
$\sigma$ model on $\PP$
there is a ``diagonal'' $\tilde T^4$ which is invariant under
the orbifold action, and is insensitive to the blow up deformations;
the symmetry comes from SCFT on
that $\tilde T^4$.

Furthermore, like the $\sigma$ model on $\PP$, the spacetime
SCFT obtained in string theory on $\MM$ appears to be unitary 
(see \egp\ and references therein).
In our description the no ghost theorem
follows from the explicit construction
of the Hilbert space of the theory. To complete the proof of unitarity
one needs to show that the set of states satisfying the 
bound \unitaritycon\ is closed under OPE.

The moduli spaces of the two CFT's agree as well. The moduli space
of $(4,4)$ superconformal $\sigma$ models on $\PP$
is twenty dimensional. Sixteen of the moduli
correspond to the metric $G_{ij}$ and antisymmetric
tensor field $B_{ij}$
on $\tilde T^4$. The remaining four moduli are certain blowing 
up modes of $\PP$. This space has singular 
subspaces fixed under various elements of $S_{kp}$. 
A standard CFT analysis shows that the only element of $S_{kp}$
whose fixed point set can be blown up by a marginal operator
is the $Z_2$ that exchanges two $\tilde T^4$'s. All other blowing
up operators are irrelevant.  

\nref\kutas{D. Kutasov, \pl{220}{1989}{153}.}%
\nref\asp{P. Aspinwall and D. Morrison, hep-th/9404151.}%
The fixed point set of the $Z_2\in S_{kp}$ is a connected $4(kp-1)$
dimensional manifold. The marginal operators that blow up
the $Z_2$ singularity have vanishing momentum along this set
(higher momentum leads to higher scaling dimension). They are
isomorphic to the blowing up modes of a {\it single} $Z_2$
singularity in CFT on $T^4/Z_2$. Therefore these blowing up
modes extend the Narain moduli space $SO(4,4)/SO(4)\times SO(4)$.
$(4,4)$ supersymmetry guarantees \refs{\kutas, \asp} that the
space is locally $SO(5,4)/SO(5)\times SO(4)$. 
The full moduli space is:
\eqn\modspce{\HH\backslash SO(5,4)/SO(5)\times SO(4)} 
$\HH$ is a discrete duality group that determines the
global structure of the moduli space. It contains the T-duality
group $SO(4,4;Z)$; a natural guess is $\HH=SO(5,4;Z)$.

In string theory on $\MM$ one finds the same twenty moduli;
the sixteen moduli $G_{ij}$, $B_{ij}$
correspond to the operators \bij\ with
$j=m=\bar m=0$. Changing these spacetime
moduli corresponds to adding to the worldsheet
Lagrangian the term $(G_{ij}+B_{ij})\int d^2z\partial Y^i\bar\partial
Y^j$. Thus, the size and shape of $\tilde T^4$
is directly related to that of $T^4$. 

The four remaining moduli are related to the chiral fields
described by the vertex operators:
\eqn\sixfr{e^{-\phi-\bar\phi}
(\psi^3-{1\over2}\gamma\psi^--{1\over2}\gamma^{-1}\psi^+)
(\bar\psi^3-{1\over2}\bar\gamma\bar\psi^--{1\over2}
\bar\gamma^{-1}\bar\psi^+)
V_{jm\bar m}V'_{jm'\bar m'}}
One can show that \sixfr\ corresponds to a chiral primary of the
$(4,4)$ superconformal symmetry \nssup, with $h=\bar h=j=\bar j$,
for all $0<j\in Z/2$. The highest components of the field \sixfr\ 
with $j=1/2$ are four singlets under $SU(2)_R\times SU(2)_L$ which are 
truly marginal in spacetime. They are described on the worldsheet
by RR vertex operators. 

The fact that the moduli space of string theory on $\MM$ is 
given by \modspce\ can be understood by noting that type 
II string theory on $T^4$ (or M-theory on $T^5$) has the 
moduli space of vacua
\eqn\modspace{SO(5,5;Z)\backslash SO(5,5)/SO(5)\times SO(5)} 
Compactifying the remaining six dimensions on $AdS_3\times
S^3$ gives a mass to five of the twenty five scalars parametrizing
\modspace\ and restricts the moduli space to \modspce. The discrete
duality group $SO(5,5;Z)$ is reduced as well. Thinking of $\MM$
as the near horizon geometry of a system of $NS5$-branes and fundamental
strings (as in section {\it 4.1}), the U-duality group is the
subgroup of $SO(5,5;Z)$ that leaves the $k$ fivebranes and $p$
strings invariant. This clearly includes the T-duality group
$SO(4,4;Z)$. It is possible that the discrete symmetry of the near
horizon theory is larger, as mentioned above.

The connection between the $T^4$ and $\tilde T^4$  parameters 
can be made more precise as follows. As an example,
take $T^4$ to be a square torus with sides $R$
and volume $v=R^4$, with $1\ll v\ll p/k$ such that
the $T^4$ is large but the description
of section 3 is still weakly coupled.
To calculate the volume of $\tilde T^4$
we would like to consider string states \physferm\ which
carry momentum along the $T^4$. These are states of the
general form $\exp(-\phi-\bar\phi)\exp(i\vec p\cdot \vec Y)
W_NV_{jm\bar m}$, with $W_N$ an operator constructed
out of the non-zero modes of the worldsheet fields as in
\physferm. The components of $\vec p$ are quantized in
integer multiples of $1/R$. We would like to compute
the spacetime scaling dimensions of the corresponding
operators and, in particular, the spacing
between subsequent momentum modes.

Substituting into \scdim\ we have:
\eqn\dimss{N-{j(j+1)\over k}+{1\over 2}|\vec p|^2={1\over2}}
Solving for $j$ we find:
\eqn\jform{j={1\over2}\left(-1+
\sqrt{1+4k(N-{1\over2})+2k|\vec p|^2}\right)}
We are interested in the dependence of the spectrum on
$\vec p$ for small $|\vec p|$. To leading order we have
\eqn\expandj{j=-{1\over2}+{1\over2}\sqrt{1+4k(N-{1\over2})}
+{1\over2}{k|\vec p|^2\over \sqrt{1+4k(N-{1\over2})}}+\cdots}
Thus the spacetime scaling dimension is proportional
to $|\vec p|^2$, and therefore the volume of $\tilde T^4$
is proportional to $v$. The precise coefficient of proportionality
depends on $N$. This is probably due to the presence of the
blowing up modes and requires more study\foot{It is interesting
to note that if we take $N$ to have the largest value
compatible with the unitarity bound \unitaritycon,
$j(\vec p=0)=(k-1)/2$, we find from \expandj\ that
$j(\vec p)=j(0)+|\vec p|^2/2+\cdots$ which seems to imply
that $\tilde T^4$ has the same volume as $T^4$.}. 

We see that the volume $v$ discussed in section {\it 4.1}
is indeed the modulus controlling
the volume of the torus $\tilde T^4$ \targets.
This identification was made in the region of validity
of the description of section 3, $v\ll p/k$. As discussed
in the previous subsections, when $v$ grows and eventually
becomes much larger than $p/k$, the description of the system
given in section 3 becomes strongly coupled and we have to
pass to the $D$ picture of subsection {\it 4.2}. There,
the volume of $T^4$, $\hat v$, is fixed at $p/k$ and the
parameter $v$ corresponds to another modulus of the SCFT on
$\PP$, perhaps one of the four blowing up modes.

Next we turn to the chiral fields of both theories.
Consider first the $\sigma$ model on $\PP$.
Denote the $\sigma$ model fields by $Z^i_A$,
with $i=1,\cdots, 4$ a vector index in the $\tilde T^4$,
and $A=1,\cdots, kp$.
The (left and right moving) fermion superpartners
of $Z^i_A$ will be denoted by $\Psi^{a\alpha}_A$,
$\bar\Psi^{b\beta}_A$, respectively, where $a$, $b$
are spinor indices of the $SO(4)$ acting on the
$\tilde T^4$ and $\alpha$, $\beta$ are spinor indices
in $SU(2)$. The orbifold in \targets\ acts on the index
$A$; the SCFT has an untwisted sector and various twisted
sectors. We will focus on the untwisted sector; the twisted
sectors can be discussed analogously.

The basic chiral operators of dimension $(h,\bar h)=
(1/2,0)$ and $(0,1/2)$ in the untwisted sector
are $\sum_A\Psi^{a\alpha}_A$,
$\sum_A\bar\Psi^{b\beta}_A$. The upper components of these
operators are $\sum_A \partial Z^i_A$ and $\sum_A
\bar\partial Z^i_A$, respectively. These chiral superfields live in
the decoupled $\tilde T^4$ sector and generate the $U(1)^4$
affine Lie superalgebra which we have identified
in the string context before.

The first non-trivial chiral operators have dimension
$(1/2, 1/2)$, and are given by\foot{Here and below a
sum over repeated indices is implied.}:
\eqn\chihalf{\Psi^{a\alpha}_A\bar\Psi^{b\beta}_A}
The operators \chihalf\ have spin $(1/2, 1/2)$
under $SU(2)_L\times SU(2)_R$ and transform as $({\bf \half}, {\bf
\half})$ under $SO(4)$.
More generally, we can define chiral operators
with $(h,\bar h)=(l/2,l/2)$:
\eqn\chil{\Psi^{a\alpha_1}_{A_1}
\Psi^{a_2\alpha_2}_{A_2}
\Psi^{a_3\alpha_3}_{A_3}\cdots
\Psi^{a_l\alpha_l}_{A_l}
\bar\Psi^{b\beta_1}_{A_1}
\bar\Psi^{a_3\beta_2}_{A_2}
\bar\Psi^{a_4\beta_3}_{A_3}\cdots
\bar\Psi^{a_2\beta_l}_{A_l}}
symmetrized over $(\alpha_1,\cdots, \alpha_l)$
and $(\beta_1,\cdots,\beta_l)$.
The operators \chil\ have $SU(2)_L\times
SU(2)_R$ spin $(l/2,l/2)$. We have summed
over $a_2, \cdots, a_l$ in \chil\ since
it is sufficient to identify operators
corresponding to low representations of $SO(4)$
in the SCFT on $\PP$ and string theory
on $\MM$. Higher representations of $SO(4)$
can then be obtained by acting with the affine
$U(1)^4$ Lie superalgebra, which has been identified
in both theories. Note also that, due to the
Fermi statistics of the $\Psi$'s,
one must have $l\le kp$.

The corresponding chiral operators
in string theory on $\MM$
are the lowest components
of superfields whose highest components are the scalars
with $(-1,-1)$ picture vertex operators (see \bij)
\eqn\scmatch{e^{-\phi-\bar\phi}
\lambda^i\bar\lambda^{\bar i}V_{jm\bar m}V'_{jm'\bar m'}.}
These fields
have $SU(2)_R\times SU(2)_L$
spin $(\tilde j,\tilde j)$ and spacetime scaling dimensions
$(h,\bar h)=(\tilde j,\tilde j)$, with $\tilde j\equiv j+
{1\over2}$. These states are in one to one
correspondence with the chiral primaries \chil, with
$j=l/2$. Note that while
$l$ in \chil\ is bounded by $kp$, the string construction
only gives rise to operators
with $l\le k-2$, since unitarity of the worldsheet
$SU(2)$ affine Lie algebra requires $j\le (k-2)/2$.
The remaining 
states are supposed to arise from multiparticle states
or bound states at threshold; it would be nice to understand
precisely how that happens. 

In the orbifold limit the $\sigma$ model on the space
$\PP$ has a large chiral algebra of operators with
$L_0=m\in Z/2$, $\bar L_0=0$.
For example, one can consider products of
the $N=4$ superconformal generators
of each $\tilde T^4$ in \targets, symmetrized to impose
the permutation symmetry \vaf. One can ask what happens to these
states with scaling dimension $(h,\bar h)=(m,0)$ when one turns on
the blowing up moduli $\alpha$. Typically, one expects the
dimensions $h$ and $\bar h$ to shift, while preserving
the spin $h-\bar h$. For small $\alpha$, the resulting $\bar h(\alpha)$
will be small and, similarly, $h(\alpha)$ will be approximately
equal to its orbifold limit value.

We have seen that operators with spin larger than
two correspond in string
theory on $\MM$ to massive string modes, such as \physnew. 
If we take $k$ to be large to make supergravity reliable, 
operators with non-zero spin 
(on the worldsheet and in spacetime) have $\bar N\ge 1$ and 
therefore due to \scdim\ large $j$ and scaling dimension
in spacetime, $h,\bar h\sim \sqrt{k\bar N}$, \levmatch.
Therefore, the spacetime SCFT obtained
{}from string theory on $\MM$ has the peculiar property that
for large $k$ and $p$ states\foot{Of course, here we are referring
to single particle states. Multi-particle states with spin
higher than two but scaling dimensions much smaller than $\sqrt k$
can and do exist.} with $2< |L_0-\bar L_0|\ll \sqrt k$ must
have $L_0, \bar L_0 \ge\sqrt k$. In particular, in the language
of SCFT on $\PP$, operators which in the orbifold limit have
scaling dimensions
$(h,\bar h)=(m,0)$ with $1\ll m\ll \sqrt k$ have after
the blow up $h,\bar h\gg m$, but $h-\bar h=m$.
This means that the deformation from
the orbifold limit is {\it large}. This is
consistent with expectations based on six dimensional
supergravity in which all the light states have small
spin \vaf. The remaining states have large dimension $\ge \sqrt k$
and hence are stringy in nature.

\subsec{BTZ Black Holes and Fundamental String States}

\nref\coushen{O. Coussaert and M. Henneaux,
hep-th/9310194, \prl{72}{1994}{183}.}
Since string theory on $\MM$
reduces to Einstein gravity in the low
energy limit, we know that it
should contain BTZ black holes \btz\
which are parametrized by their mass
$M$ and angular momentum $J$.
The Lorentzian signature black hole metric is given by:
\eqn\th{\eqalign{
ds^2=&-N^2dt^2+N^{-2}dr^2
+r^2(N^\phi dt+d\phi)^2\cr
N^2=&(r/l)^2-8l_pM+(4l_pJ/r)^2\cr
N^\phi=&-4l_pJ/r^2\cr
}}
One can think of the black holes \th\
as excitations of the vacuum described
by the $M=J=0$ black hole, \ie\ the
Ramond vacuum of the spacetime SCFT \refs{\btz, \strom}.
The mass and spin of the black holes are given
in terms of the Virasoro generators in the Ramond
sector $L_0$, $\bar L_0$ by:
\eqn\massspin{Ml=L_0+\bar L_0,\qquad  J=L_0-\bar L_0}
where we have defined $L_0$ such that
it vanishes on the Ramond vacuum (by subtracting
$c/24=kp/4$ from $\tilde L_0$ \mapops).
Some of the solutions \th, namely
those with $J=\pm Ml$,  preserve
half of the supersymmetries of the
$M=0$ vacuum \coushen.
In the spacetime SCFT
these are states with either $L_0=0$
or $\bar L_0=0$.

\nref\dasmat{S. Das and S. Mathur, hep-th/9601152.}%
\nref\malsus{J. Maldacena and L. Susskind,
hep-th/9604042.}%
Note that the correspondence
\massspin\ means that from the spacetime gravity
point of view the lowest lying BTZ black holes are very
light. Since the low lying states in the spacetime CFT
have $L_0\sim 1$, the lowest mass BTZ black holes
have $M\sim 1/l$ and are much lighter than the natural
mass scale of the theory $1/l_p$. In fact, comparing
\cspacetime\ and \brhencent\ we see that $M\sim
1/(l_p kp)$. Note also that the 
scales $l$ and $l_p$ of our system depend differently
on the parameters $k,p,v$ of the model in the two different
regions corresponding to weak $NS$ and $D$ coupling discussed
in the previous subsections:
\eqn\diffreg{\eqalign{
D:\;\;\;\;\; l=&2\pi l_s\left({kp\over v}\right)^{1\over4},
\;\;\;l_p={\pi\over2}l_s(kp)^{-{3\over4}}v^{-{1\over4}}\cr
NS:\;\;\;\;\; l=&l_s\sqrt{k}, \;\;\;l_p={l_s\over 4p\sqrt k}\cr}}

The Bekenstein-Hawking entropy of BTZ black holes
with mass $M$ and angular momentum $J$ has the
usual form in terms of the area $A$ of the event
horizon:
\eqn\ti{S={A\over 4l_p}=\pi\sqrt{l(lM+J)\over 2G_3}
+\pi\sqrt{l(lM-J)\over 2G_3}=2\pi\sqrt{kpL_0}+2\pi\sqrt{kp\bar L_0}}
How can one describe BTZ black holes in the framework of
our previous analysis? The states we are looking for should have
finite masses \massspin\ in the weak coupling limit $p\gg 1$,
$M\sim L_0/l_s\sqrt{k}$. Thus, we would like to identify them
with fundamental string states. By the analysis of the previous
sections, we can associate to every perturbative string
state a value of $L_0$ and $\bar L_0$ and, therefore,
\massspin\ a mass and spin.

Of course, string states should only
be thought of as black holes if their horizon area
is larger than their size, which is of order $l_s$.
In fact, while one can construct large black holes
with $A\gg l_s,l_p$
{}from multi-particle perturbative string states
at weak string coupling, the perturbative description
is {\it not} valid for such black holes. 
Indeed, substituting
\weakcoup\ in \ti\ one finds that $A\sim l_s\sqrt{L_0/p}$; thus
large black holes necessarily have $L_0\gg p$.
The corresponding energies are of order $1/g_6^2$
(with the six dimensional string coupling $g_6^2\propto k/p$),
and the perturbative string picture is not expected to be
reliable at such high energies. It is nevertheless possible that one can
use the large symmetry of this system to obtain useful
information about the physics of large black holes.

BTZ black holes with $Ml=\pm J$
(\ie\ vanishing $L_0$ or $\bar L_0$ \massspin)
correspond in string theory to
multi-particle states constructed out of the chiral algebra
modes $L_{-n}$, $T^a_{-n}$, $\alpha_{-n}^i$, \etc.
Most of the black holes correspond to massive
string states, have non-zero $L_0, \bar L_0$
and break the supersymmetry completely.

\nref\cardy{J. Cardy, \np{270}{1986}{186}.}%
\nref\carlip{S. Carlip, hep-th/9806026.}%
\nref\cnd{O. Coussaert, M. Henneaux and P. van Driel,
gr-qc/9506019, \cqg{12}{1995}{2961}.}%
The above construction of BTZ black holes in string theory
on $\MM$ allows one to compute the Bekenstein-Hawking
entropy of these objects. Since BTZ black holes are
described in our framework as multiparticle states
in the spacetime SCFT, we can apply the standard result
{}from CFT \cardy, to compute the entropy $S$:
\eqn\entbh{S=2\pi\sqrt{cL_0\over 6}+2\pi\sqrt{c\bar L_0\over 6}}
The central charge of the spacetime
theory is $c=6kp$ \cspacetime; thus \entbh\ agrees with the form \ti\
we found by using the area formula before.

\nref\birm{D. Birmingham, I. Sachs and S. Sen, hep-th/9801019,
\pl{424}{1998}{275}.}%
This argument is due to Strominger \strom\ (see also \birm). Our 
analysis supplements that of \strom\ in two respects:
\item{(a)} The formula \entbh\ only applies to CFT's for
which the lowest dimension operator has $h=\bar h=0$.
In the context of gravity on $AdS_3$ it was applied to the
CFT living on the boundary of $AdS_3$, but it was not clear
whether this condition applies (see \refs{\emil, \carlip}
for recent discussions).
In fact it has been argued that the boundary (S)CFT is a 
(super-)Liouville theory \cnd, for which $c$
should be replaced \refs{\liouvilles,\ks}\ by
$c_{\rm eff}=1$ in \entbh. Our
string theory on $\MM$ is unitary and its lowest 
dimension operator has $h=0$ (the identity operator). 
Therefore, the conditions for applying \entbh\ are satisfied
here, at least for weak coupling (\ie\ for large enough $p$).
\item{(b)} We showed that the states contributing to the
density of states \entbh\
are fundamental string states,
most of which are furthermore massive; therefore, one cannot
reduce to supergravity without losing the microscopic interpretation
of \ti.

\bigskip
\noindent{\bf Acknowledgements:}
We thank S. Elitzur, J. Harvey, E. Martinec, E. Rabinovici, 
A. Strominger, E. Witten, and especially A. Schwimmer for 
discussions. The work of A.G. is supported in part by BSF 
-- American-Israel Bi-National Science Foundation and by the 
Israel Academy of Sciences and Humanities -- Centers of 
Excellence Program. A.G. thanks the Einstein Center at the Weizmann
Institute for partial support. The work of D.K. is supported in part 
by DOE grant \#DE-FG02-90ER40560. The work of N.S. is supported by 
DOE grant \#DE-FG02-90ER40542.

\appendix{A}{The Geometry of Lorentzian $AdS_3$}

The Lorentzian signature version of $AdS_3$ is obtained by
analytically continuing the Euclidean version described by
eq. \defeq. The inequivalent continuations correspond to replacing
$X_3=iX_0$ and $X_1=iX_0$ in \defeq.  Clearly there should not be any
difference between them.  The first corresponds to setting $\tau=it$
in \tthetacoo\ and \metricttheta.  It leads to\foot{We set $l=1$ in
this appendix.}
\eqn\metrictthetam{ds^2={1 \over 1+r^2}dr^2-(1+r^2)dt^2+r^2d\theta^2.}
The second corresponds to treating $\gamma$ and $\bar \gamma $ as two
independent real coordinates and letting $u=e^{-\phi}$ in \uycoo\ be
both positive and negative (now $u=0$ is not a boundary of the
space).  The metric is
\eqn\metricupypm{ds^2={1 \over u^2}du^2+u^2d\gamma d\bar \gamma.}
Because of the reality properties of $\gamma$ and $\bar \gamma$, we
can no longer use \changeva\ to relate them.  Instead, these two
coordinate systems are related by the transformations
\eqn\primecoom{\eqalign{
&u=\sqrt{1+r^2}\cos t + r\cos \theta \cr
&\gamma={\sqrt{1+r^2}\sin t+ r\sin \theta \over \sqrt{1+r^2}\cos t +
r\cos \theta} \cr
&\bar \gamma={-\sqrt{1+r^2}\sin t+ r\sin \theta  \over \sqrt{1+r^2}
\cos t + r \cos \theta} \cr}}
and the inverse map
\eqn\primecoomi{\eqalign{
&r^2={[u^2(\gamma \bar \gamma-1)+1]^2 \over 4u^2} + {(\gamma+\bar
\gamma)^2u^2\over 4} =
{[u^2(\gamma \bar \gamma+1)+1]^2 \over 4u^2} +{(\gamma-\bar
\gamma)^2u^2 \over 4} -1 \cr
&\sin t={u(\gamma - \bar \gamma)\over 2\sqrt{1+r^2}}\cr
&\sin\theta={u(\gamma + \bar \gamma) \over 2r} \cr}}
where in the last two expressions we use $r$ from the first.  Note
that the expression for $r^2$ is always non-negative and, therefore, the
square root can be taken.  Similarly, the two ways of writing $r^2$
guarantee that $\sin t$ and $\sin \theta$ are in $[-1,1]$.  This shows
that for $t\in [0,2\pi)$ and $u\in (-\infty,\infty)$ the change of
variables \primecoom, \primecoomi\ is one to one.

The relation to the $SL(2,R)$ group manifold
is obtained by parametrizing it by the Gauss decomposition
\eqn\defg{g=\left(\matrix{
1&\bar\gamma \cr
0&1\cr}\right)
\left(\matrix{
{1\over u} &0\cr
0&u\cr}\right)\left(\matrix{
1&0\cr
\gamma &1\cr}\right)=
\left(\matrix{
\gamma \bar\gamma  u +{1 \over u} & \bar\gamma  u\cr
\gamma  u & u \cr} \right)}
where $\gamma$ and $\bar \gamma$ are two independent real numbers.
The metric on the group is
\eqn\groupmet{ds^2=\half \Tr (g^{-1}dg)^2 = {1 \over u^2}(du)^2 +
u^2 d\gamma d\bar\gamma }
which is readily identified with \metricupypm.
Here it is clear that the $SL(2,R)$ group manifold is described by
$u\in (-\infty,\infty)$.

For physics we would like the time $t$ to be non-compact.  Therefore,
although in the original space there is a closed time-like curve
corresponding to $t\in [0,2\pi)$, we must consider the infinite cover
of this space with $t \in (-\infty,\infty)$.  Now we see that the
transformations \primecoom, \primecoomi\ are no longer one to one.
Points which differ by $t \rightarrow t+2\pi$ are mapped to the same
$u, \gamma, \bar \gamma $.

\appendix{B}{Twisted Strings on $\MM$}

In section 3 we discussed superstring theory on the manifold
$\MM$ \one. Our construction of spacetime supercharges did not
follow the usual route of identifying a global $N=2$ superconformal
symmetry on the worldsheet and using the $U(1)_R$ current
inside that $N=2$ algebra to construct the spacetime
supercharges in the standard way. In this appendix we will
describe the string vacuum that is obtained by following
the usual path. This vacuum seems to be related
to the Ramond vacuum of the spacetime SCFT discussed above by
twisting the spacetime supersymmetry and treating it as a BRST
charge. This interpretation of the construction below is conjectural
and requires a much better understanding.

\lref\rrocek{I. T. Ivanov, B. Kim and M. Rocek,
\pl{343}{1995}{133}}%
To enhance the $N=1$ superconformal algebra \TG\ to $N=2$
we must find a $U(1)_R$ current, $J_{N=2}$, which is part of the $N=2$
algebra. In our case this $U(1)_R$ current can be
chosen to be\foot{In this appendix we normalize
$\psi^+\psi^-=i\psi^1\psi^2$, \etc. This differs by a factor
of two from the normalization used in the text.} \rrocek:
\eqn\Jz{J_{N=2}(z)=
-{2(k+2)\over k^2}\psi^+\psi^-
+{2(k-2)\over k^2}\chi^+\chi^-
+{2\over k}j^3-{2\over k}k^3
+{2\over k}\psi^3\chi^3
+i\lambda^1\lambda^2+i\lambda^3\lambda^4}
\Jz\ is unique up to global symmetries\foot{Assuming
that we do not want to mix the $SL(2,R)\times SU(2)$ and
$T^4$ parts of the theory.}.
It can be obtained by decomposing
\eqn\decomp{\MM\simeq {SL(2,R)\over U(1)}\times
{SU(2)\over U(1)}\times U(1)^2\times U(1)^4}
Each of the factors in \decomp\ has a natural
complex structure; \Jz\ (as well as the other
$N=1$ superconformal generators \TG) can be written as a sum
of the corresponding currents:
\eqn\ddeecc{
\eqalign{
J_{N=2}=&J^{(1)}_{N=2}+J^{(2)}_{N=2}+J^{(3)}_{N=2}+J^{(4)}_{N=2} \cr
J^{(1)}_{N=2}=&-{2\over k}\left(\psi^+\psi^--J^3\right)\cr
J^{(2)}_{N=2}=&{2\over k}\left(\chi^+\chi^--K^3\right)\cr
J^{(3)}_{N=2}=&{2\over k}\psi^3\chi^3\cr
J^{(4)}_{N=2}=&i\lambda^1\lambda^2+i\lambda^3\lambda^4\cr
}}
where $J^3$ and $K^3$ are the total
$SL(2,R)$ and $SU(2)$ currents defined
in \Jp.
Note that
the choice of the complex structure \Jz, \ddeecc\
breaks $SL(2,R)\times SU(2) \to U(1)\times U(1)$.
It is also useful for future purposes
to note that the currents $J^3$ and $K^3$
have non-singular OPE's with the $U(1)_R$ currents
$J^{(i)}_{N=2}$ \ddeecc.

$T$, $G$, $J_{N=2}$, \TG, \Jz\ generate together
an $N=2$ superconformal algebra.
The supercurrent $G$ \TG\ splits into
two parts, $G=G^++G^-$ with charges
$\pm1$ under $J$.
To construct the spacetime supercharges
\fms, we bosonize the $U(1)_R$ current
\Jz:
\eqn\JiH{J_{N=2}=i\partial(H+H_1+H_2+H_3)}
where $H,H_{1,2,3}$ are chiral scalar fields obeying
\eqn\iH{\eqalign{
i\partial H=&
{2\over k}\left(-\psi^+\psi^-
+\chi^+\chi^-
+J^3-K^3\right)\cr
i\partial H_1=&
{2\over k}\psi^3\chi^3, \qquad
\partial H_2=\lambda^1\lambda^2, \qquad
\partial H_3=\lambda^3\lambda^4\cr
}}
The fields $H_I$, $I=1,2,3$ are normalized canonically,
$\langle H_I(z)H_J(w)\rangle=-\delta_{IJ}\log(z-w)$,
while $\langle H(z)H(w)\rangle=-2\log(z-w)$.
The spacetime supercharges are the zero-modes
$Q^{\pm}_{\alpha}$, $\bar Q^{\pm}_{\bar\alpha}$
of the eight mutually local BRST invariant spin-fields:
\eqn\qpm{
\eqalign{
S^{\pm}_{\alpha}=&e^{-{\phi\over 2}\pm
{i\over 2}H+{i\over2}H_1} S_\alpha, \qquad Q^{\pm}_{\alpha}=
\oint dz S^{\pm}_{\alpha}\cr
\bar S^{\pm}_{\bar\alpha}=&e^{-{\phi\over 2}\pm
{i\over 2}H-{i\over2}H_1} S_{\bar\alpha}, \qquad
\bar Q^{\pm}_{\bar \alpha}=
\oint dz \bar S^{\pm}_{\bar\alpha}\cr
}}
$S_\alpha, S_{\bar\alpha}$ are spinors
in the ${\bf 2}, {\bf \bar2}$ of the $SO(4)$ symmetry
acting on the fermions
$\lambda^i$:
\eqn\salpha{S_{\alpha}=e^{\pm{i\over 2}(H_2+H_3)}
, \qquad S_{\bar\alpha}=e^{\pm{i\over 2}(H_2-H_3)}
}
The spacetime superalgebra is
\eqn\QQP{
\eqalign{
\{Q^+_\alpha,Q^-_\beta\}&=\delta_{\alpha\beta}(J^3-K^3)\cr
\{\bar Q^+_{\bar\alpha},\bar Q^-_{\bar\beta}\}&=
\delta_{\bar\alpha\bar\beta}(J^3+K^3)\cr
\{Q^-_\alpha,\bar Q^+_{\bar\beta}\}&=
\{Q^+_\alpha,\bar Q^-_{\bar\beta}\}=
\gamma^i_{\alpha \bar\beta}P_i\cr
}}
All other (anti-) commutators vanish.
$P_i$ is the four-vector of momenta along the
$T^4$, and $J^3,K^3$ are the zero-modes of the
total CSA currents \Jp.

The first line of the
superalgebra \QQP\ looks similar to the
Ramond sector superalgebra \nssup\ in the zero
mode sector. We have four supercharges $Q^\pm_\alpha$
which one can attempt to identify\foot{The symmetry structure
implies that the index $\pm$ on $Q^\pm_\alpha$ corresponds to
the index $i$ on $Q$, $\bar Q$ in \nssup, while $\alpha=+$ and
$\alpha=-$ correspond to $Q$ and $\bar Q$ in \nssup.}
with $Q^i_0$, $\bar Q^j_0$ in \nssup.
$J^3-K^3$ on the r.h.s. of the first line of \QQP\
is then interpreted as $L_0$, which is just the
way the Ramond sector $L_0$ is expected to be related
to the NS generators $L_0=-J^3$ and
$T^3_0=K^3$, as in the last line of
eq. \mapops. It is important for the above interpretation
of $Q^\pm_\alpha$ and $J^3-K^3$ that the supercharges
commute with the bosonic generators $J^3$, $K^3$.

The second line of \QQP\ appears to describe another copy
of the same structure as the first line, with four more
supercharges, $\bar Q^\pm_{\bar\alpha}$ which square to
the bosonic generator $J^3+K^3$. At first sight it
looks like we should interpret this as the
Ramond superalgebra corresponding to the other
chirality in spacetime, but this appears to be inconsistent
for the following reasons:
\item{(a)} The third line of \QQP\ would then say that left and right
moving supercharges in spacetime have non-zero anticommutators.
Such terms indeed arise in two dimensional $(4,4)$ supersymmetric
theories, \eg\ when the latter are obtained by compactifying
$N=1$ supersymmetric six dimensional theories; the charges $P_i$
in \QQP\ correspond to central charges in the superalgebra. However,
such central charges are inconsistent with the {\it conformal} symmetry
that string theory on $AdS_3$ is supposed to possess.

\item{(b)} We saw in the text that in string theory on $AdS_3$ the
worldsheet and spacetime chiralities are related. It would be
strange to get both left and right movers in the spacetime SCFT
{}from the same chirality on the worldsheet. A related problem
is that if this had happened, we would have had trouble interpreting
the eight additional supercharges and translation generators
$\bar J^3$, $\bar K^3$ arising from the other worldsheet chirality.

\noindent
A clue towards the correct interpretation of the
vacuum in question comes from noticing that while
the superalgebra \QQP\ looks symmetric
under interchange of $J^3-K^3$ and $J^3+K^3$,
this is misleading. The choice of the complex
structure \Jz, which contains the term \ddeecc\
$2(J^3-K^3)/k$, picks one over the other.
This leads to an asymmetry of the spectrum
of excitations, which must satisfy the GSO
projection (\ie\ have integer $U(1)$ charges
under \Jz). There is another vacuum in which we flip
the relative sign between $J^3$ and $K^3$ in \Jz\
and in which the roles of $J^3\pm K^3$ are reversed.

Therefore, we would like to propose that what we are
actually describing here is the two possible Ramond vacua
corresponding to different twists \spectflow\
of the NS vacuum of section 3. Of course, we cannot be
describing both vacua of the theory at the same time.
Thus, to make sense of the theory we are instructed to
do the following. If the $U(1)_R$ current $J_{N=2}$
contains the combination $J^3-K^3$ as in \Jz, we define
the theory by restricting physical states to the cohomology
of the operators $Q^\pm_\alpha$:
\eqn\physcond{Q^\pm_\alpha|{\rm phys}\rangle=0,\;\;\;
|{\rm phys}\rangle\sim |{\rm phys}\rangle+Q|{\rm anything}\rangle}
Because of \QQP\ all such states
have $J^3-K^3=0$. We then interpret the four remaining
supercharges $\bar Q^\pm_{\bar\alpha}$ together with $J^3+K^3$
as forming the zero mode sector of the left moving $N=4$ superconformal
algebra \nssup. Obviously, if $J_{N=2}$ contains the combination
$J^3+K^3$, we reverse the roles of $Q$ and $\bar Q$ and of
$J^3\pm K^3$.

Note that this procedure resolves the difficulties
mentioned above. The non-zero anticommutators in
the third line of \QQP\ are no longer relevant since
all the states are killed by $Q$. In fact, \QQP\ implies
that physical states have $P^i=0$. The reason is that
we expect the spacetime theory to be unitary, and positivity
of the norm of physical states implies that they satisfy
\eqn\genpostv{(J^3)^2\geq (K^3)^2+|\vec P|^2}
Since $J^3=K^3$, one must have $P^i=0$.
The doubling of the superalgebra compared
to what one expects in the Ramond sector is avoided
by imposing the condition \physcond\ on physical states.
And, the correlation between the worldsheet and spacetime
chirality is restored: left moving symmetries on the worldsheet
give left moving symmetries in spacetime, and vice-versa.

To summarize, string theory on $\MM$ with the GSO projection
related to the $U(1)_R$ current \Jz\ is non-unitary
(see below). One can restrict to a unitary sub-sector
by restricting to the cohomology of $Q^\pm_\alpha$ \physcond.
This unitary sub-sector describes in spacetime the
{\it chiral ring} or zero mode sector of the $D1/D5$
SCFT, obtained by restricting to states satisfying
$L_0=T^3_0$ (and similarly $\bar L_0=\bar T^3_0$).
The projection \physcond\ is an analog of the twist
one does in $N=2$ SUSY theories in two dimensions, which
leads to a topological $N=2$ theory whose physical states
are in one to one correspondence with the chiral ring
of the original theory.

To verify this interpretation
we next turn to the spectrum of excitations of the
theory. Due to the spacetime supersymmetry we can
restrict our attention to the Neveu-Schwarz
sector of the worldsheet theory (\ie\ concentrate
on spacetime bosons); we will study
the spectrum in the $-1$ picture of \fms\ and
continue to work chirally. As discussed above,
we will also set the momenta along the
$T^4$, $P_i$, to zero.

Before the projection \physcond\
the system appears to contain tachyons,
analogs of those that exist in
fermionic string theory in flat space.
They are described by the vertex operators
(we consider only the holomorphic part of the vertex
operators):
\eqn\tach{T_{jmj'm'}=e^{-\phi}V_{jm}V'_{j'm'}}
where, as in the text, $V_{jm}$ is a primary operator
of $SL(2)$ affine Lie algebra with
quadratic Casimir $-j(j+1)$ and $J^3=m$;
$V'_{j'm'}$ is an $SU(2)$ primary with similar notation.
Unitarity restricts the allowed
values of $j',m'$,
\eqn\jktwo{j'\leq {k\over2}-1, \;\;\;|m'|\le j'}
The mass shell condition on $T$ is in this case
\eqn\massshell{
{j'(j'+1)\over k}-{j(j+1)\over k}={1\over2}
}
The GSO projection provides a constraint on $m$, $m'$:
\eqn\gsop{{m'-m\over k}\in Z+{1\over2}}
Thus, these states disappear from the spectrum
in the flat space limit $k\to\infty$.
Many of the solutions of \tach\ --
\gsop\ have complex $j=-(1/2)+i\lambda$ and, therefore,
correspond to tachyons; an example is
states with $j'=m'=0$, $m=(2n+1)k/2$ for
integer $n$. Therefore, before the projection
\physcond, the string vacuum in question is
unstable and the spacetime dynamics in it
is not unitary. The projection \physcond\
eliminates the tachyons \tach\ since due to
\gsop\ none of these modes satisfy $J^3=K^3$.

The low lying states of the theory are ``transverse photons.''
As usual there are eight physical polarizations,
four along the $T^4$ and four living in $SL(2,R)
\times SU(2)$. The photons polarized along the
$T^4$ are described by the vertex operators
\eqn\lVV{W^i_{jm}=e^{-\phi}\lambda^iV_{jm}V'_{jm}}
where we set $j=j'$ so that the total scaling
dimension of \lVV\ is one, and $m=m'$ to enforce
\physcond, $J^3=K^3$. The four remaining light
transverse photons are described by
vertex operators of the form
\eqn\morephot{\eqalign{
W=e^{-\phi}\big[&a_+\psi^+ V_{j m-1}V'_{jm}
+a_-\psi^-V_{j m+1}V'_{jm}+\cr
&b_+\chi^+ V_{j m}V'_{jm-1}
+b_-\chi^- V_{j m}V'_{jm+1}+
(c_3\psi^3+d_3\chi^3)V_{jm}V'_{jm}\big]\cr}
}
BRST invariance and the freedom to add
BRST commutators to $W$ imply in the usual
way that four of the six independent polarizations
in \morephot\ are physical.

In addition to the photons \lVV, \morephot,
the spectrum also includes eight towers
of oscillator states obtained \eg\ by replacing
$\lambda^i$ in eq. \lVV\ by an $N=1$
superconformal primary
with scaling dimension $h=N+1/2$. This leads to the
standard exponential density of states at level $N$.
The $SL(2,R)$ and $SU(2)$ Casimirs obey
in this case the relation:
\eqn\hmass{{j(j+1)\over k}={j'(j'+1)\over k}+N}
Of course, one still has to impose the projection
$J^3=K^3$ and the bound \unitaritycon.
All operators \lVV\ -- \hmass\ satisfy $L_0=T^3_0$
and, therefore, are chiral. Their form and degeneracies
are in agreement with the discussion of chiral operators
in section 3.

Note that unlike the discussion in the text (sections 2, 3), 
here we {\it cannot} find an infinite super-Virasoro
algebra in spacetime. The reason is that, as discussed above,
this vacuum describes the topological dynamics of the
chiral ring, on which the infinite dimensional algebra
does not act.

\listrefs
\end